# Coordinated Allocation of Radio Resources to Wi-Fi and Cellular Technologies in Shared Unlicensed Frequencies

DAVID CANDAL-VENTUREIRA, FRANCISCO JAVIER GONZÁLEZ-CASTAÑO, FELIPE GIL-CASTIÑEIRA, AND PABLO FONDO-FERREIRO
Information Technologies Group, AtlanTTic, University of Vigo, EE Telecomunicación, 36310 Vigo, Spain

Corresponding author: Felipe Gil-Castiñeira (xil@gti.uvigo.es)

This work was supported in part by the Ministerio de Ciencia e Innovación, Spain, under Grant PID2020-116329GB-C21, in part by the Xunta de Galicia under Grant GRC2018/053, and in part by the ''la Caixa'' Foundation (ID 100010434) Fellowship under Grant LCF/BQ/ES18/11670020.

**ABSTRACT** Wireless connectivity is essential for industrial production processes and workflow management. Moreover, the connectivity requirements of industrial devices, which are usually long-term investments, are diverse and require different radio interfaces. In this regard, the 3GPP has studied how to support heterogeneous radio access technologies (RATs) such as Wi-Fi and unlicensed cellular technologies in 5G core networks. In some cases, these technologies coexist in the same spectrum. Dynamic spectrum sharing (DSS), which has already been proven to increase spectrum efficiency in licensed bands, can also be applied to this scenario. In this paper, we propose two solutions for mobile network operators (MNOs) or service providers to dynamically divide (multiplex) the radio resources of a shared channel between a Wi-Fi basic service set (BSS) and one or several carriers of scheduled wireless networks, such as cellular technologies, with a configurable level of sharing granularity. These solutions do not require modifications to the current commercial off-the-shelf (COTS) end devices. We adapt the existing IEEE 802.11 procedures to notify the Wi-Fi stations that they must share channels with different access networks. We demonstrate that our dynamic sharing proposals are also advantageous over direct coexistence and evaluate each of them quantitatively and qualitatively to determine when one or the other is preferable. The evaluation is particularized for IEEE 802.11ac and long-term evolution (LTE) license assisted access (LAA), but the solutions can be easily extended to 5G new radio-unlicensed (5G NR-U) or to any other wireless technology in which the network side schedules end device transmissions.

**INDEX TERMS** 5G mobile communications, dynamic spectrum sharing, resource sharing, unlicensed spectrum, industrial networks.

## I. INTRODUCTION

Industry is increasingly relying on wireless network technologies: 86% of manufacturing executives believe that smart manufacturing based on connected machines will be a driver of competitiveness in the near future [1]. Network performance and service level agreement (SLA) compliance are critical in these scenarios. Lower latencies, for example, may improve the productivity of automation chains, and even short downtimes can seriously cut into revenues in major ways.

The associate editor coordinating the review of this manuscript and approving it for publication was Faouzi Bouali.

Diverse wireless technologies are currently used to connect industrial machinery and workers [2], [3], and many of these technologies operate in unlicensed industrial, scientific and medical (ISM) bands. Each technology provides end devices with different performance levels in terms of throughput, latency, reliability, coverage, and mobility. Although 5G networks have been explicitly designed to support a wide range of use cases with very diverse requirements (including industrial applications with strict requirements), upgrading an entire industrial wireless network infrastructure is not feasible. Industries are conservative in that they keep their production solutions running for as long as they work [4] (as they did in the past, for example, with RS-485 wiring).

The licensed spectrum is a scarce resource that is seldom available for the deployment of private 5G networks in factories or other environments. Fortunately, the 3GPP is also considering license-exempt spectrum frequencies in the 5, 6 and 60 GHz bands [5], [6], and different trials are have been deployed. Many multinational mobile network operators (MNOs), for instance, including AT&T, T-mobile and Vodafone, are known to be investing in trials and pilots based on such technologies, and multiple commercial networks have been deployed recently [7], [8]. Some of the scenarios addressed in unlicensed bands include data offloading (based on previous work on LTE networks) and standalone operation in the license-exempt spectrum, supporting new use cases such as private cellular network deployments for industry. Furthermore, 5G networks also support non-3GPP radio access technologies, including Wi-Fi [9]. This allows operators to integrate both cellular and Wi-Fi data services in the license-exempt spectrum to enhance network performance in a cost-effective way. 5G technology will therefore allow MNOs to provide networking services in industrial facilities through heterogeneous radio access technologies (RATs). The corresponding access networks will share the same unlicensed bands. Static allocation of separate frequencies to different technologies would be far from optimal in most situations, because none of the technologies can take advantage of the others' surplus resources. A managed solution for providing dynamic spectrum sharing (DSS) in an unlicensed band (including sporadic full access) to devices of different radio technologies may achieve better overall capacity and help operators to meet SLAs.

In this paper we propose two solutions that would allow an MNO to dynamically distribute (i.e. multiplex) unlicensed radio resources, which the MNO can exclusively operate between a Wi-Fi basic service set (BSS) and a scheduled network, with different levels of resource sharing granularity. By dynamic distribution we refer to the ability to change the resource sharing ratio at any time. Our solutions rely on certain procedures of IEEE 802.11 standards, such as clear-to-send-to-self (CTS-to-self) and channel switch announcement (CSA), which we use to notify Wi-Fi devices that they must share the channel with other devices that are using different radio technologies. Of note, these solutions are natively supported by current commercial off-the-shelf (COTS) end devices. Although the proposals focus on LTE license assisted access (LAA) they can be easily extended to 5G New Radio-Unlicensed (5G NR-U), given their similar procedures, as well as to any other wireless networking technology whose transmissions are scheduled on the network side. In fact, 5G NR-U is built upon LTE LAA [10]. Summing up, our main contributions are 1) the adaptation of legacy mechanisms for dynamic sharing of heterogeneous unlicensed radio access resources in the context of 5G networks and 2) a study of the performance of these sharing methods in terms of network performance and resource granularity, particularized for IEEE 802.11ac networks using aggregate medium access control (MAC) protocol data unit (A-MPDU) aggregation, which is a common configuration in current Wi-Fi networks. As a baseline, we compare our proposed solutions with direct technology coexistence within the same channel. We evaluated our dynamic multiplexing proposals both quantitatively and qualitatively, indicating when one is preferable to the other in a particular scenario. We demonstrate that, in addition to allowing MNOs to configure the proportion of radio resources to be allocated to each technology, our proposals also outperform direct coexistence in most scenarios.

The rest of this paper is structured as follows: Section II reviews the background of this research. Section III discusses related work. Section IV describes the dynamic channel sharing proposals, which are characterized in Section V in terms of sharing granularity, channel availability and network capacity. Section VI obtains analytical results with a model for computing the throughput of a Wi-Fi 802.11ac network with A-MPDU aggregation and an LAA network, which are validated with NS-3 simulations. Section VII discusses these results and Section VIII presents our conclusions.

## II. BACKGROUND

We propose two solutions to allow dynamic and flexible allocation of spectrum resources to a scheduled network (such as 4G or 5G) and a Wi-Fi network using unlicensed frequencies, enabling more robust and flexible services in environments such as industry. In the next sections we discuss the background that has made this possible.

### A. CELLULAR DATA OFFLOADING TO LICENSE-EXEMPT BANDS

Cellular network traffic continues to grow, with a 30% compound annual growth rate predicted by 2024 [11]. Consequently, MNOs must deploy denser radio access networks (RANs) for each single cell to serve fewer users using methods such as carrier aggregation (CA), introduced in LTE Release 10 [12], which allows multiple LTE channels to be combined into a single logical carrier. User equipments (UEs) with CA capabilities can simultaneously transmit and receive data, even through multiple highly separated carriers. LTE-Advanced Pro (Release 13) can supply up to 640 MHz bandwidth with 32 carriers [13], outperforming legacy LTE channels of 20 MHz at most. This enables 25.6 Gbps peak downlink rates with $8 \times 8$ multiple-input multiple-output (MIMO) and 256 QAM modulation [14].

The licensed spectrum, however, is scarce and extremely expensive. Since MNOs rarely have over 60 MHz of licensed spectrum at their disposal for LTE channels [15], data aggregation is a very interesting option for increasing network capacity in scenarios where users demand best-effort services. There have been two typical approaches in this regard: complementing cellular communications with pre-existing technologies (e.g. Wi-Fi) and deploying LTE on license-exempt bands.

LTE-unlicensed (LTE-U) was the first LTE solution to offload traffic from licensed to license-exempt bands [16]

proposed and specified by the LTE-U Forum. It leverages the CA mechanism for aggregating licensed LTE carriers with other carriers in the 5 GHz ISM band, to which low-priority downlink data traffic is diverted. The scheduling of transmission resources (physical resource blocks or PRBs) for secondary channels is similar to that for the main channel, but carrier-sense adaptive transmission (CSAT) is used to share the medium with other services. LTE-U eNodeBs can adjust the ratio of their ''on'' and ''off'' periods according to channel usage, but any LTE-U transmission may start just after the beginning of an ''on'' period. That is, if another non LTE-U device is transmitting on the same channel, transmissions may collide unless some coordinator forbids non LTE-U ones during LTE-U ''on'' periods.

Similar to LTE-U, that is, by leveraging LTE CA capabilities, the LAA 3GPP standard for offloading traffic to license-exempt LTE carriers [17] aggregates secondary channels (with a typical bandwidth of 20 MHz) in the 5 GHz ISM band with primary carriers in the licensed spectrum. Up to 80 MHz can be aggregated into the unlicensed bands. LTE LAA employs a listen before talk (LBT) MAC procedure that, unlike LTE-U, satisfies many national regulations, especially those of the European Union. LAA evolved into enhanced license assisted access (eLAA) in LTE Release 14 [18] and into further enhanced LAA (feLAA) in LTE Release 15 [19]. The major improvements were the capabilities for transmitting uplink (UL) data traffic in license-exempt bands and autonomous uplink (AUL) transmissions in a predefined set of resources. feLAA reaches a maximum aggregated bandwidth of 100 MHz in the license-exempt bands.

MulteFire [20] is another LTE variant for the license-exempt spectrum based on LAA that allows the deployment of complete standalone base stations for private environments as an alternative to Wi-Fi networks. As with LTE-U, MulteFire specifications were developed outside 3GPP standards by an independent alliance including Qualcomm, Huawei, Ericsson and Nokia.

It is also worth mentioning that LTE Release 13 introduced a mechanism for offloading traffic to Wi-Fi networks called LTE-WLAN aggregation (LWA) [17]. MNOs that operate Wi-Fi hotspots can reduce the load on their licensed cellular carriers with this solution. However, the previous approaches for offloading traffic to LTE carriers in license-exempt bands are simpler and can be seamlessly integrated into existing LTE core networks.

Finally, the 3GPP has started supporting 5G NR in the unlicensed spectrum with the launch of 3GPP Release 16 [21], by following both a licensed assisted access and a standalone unlicensed operation, with the latter being conceptually similar to MulteFire [10].

### B. 5G NETWORK FLEXIBILITY
5G user plane functions (UPFs) [22] allow a highly granular configuration of data routes and treatment of individual data flows. This allows flexible responses to network events, even in real time.

The bandwidth part (BWP) is a very interesting 5G NR concept in this context [23]. A BWP is a group of contiguous PRBs that behave in practice as an independent channel whose control signals are exclusively transmitted within the corresponding frequencies. Consequently, a base station can support many different use cases *at different times* through, for example, a single BWP for high-bandwidth transmissions occupying the entire channel, or several smaller multiplexed BWPs to serve low-bandwidth or low-autonomy devices. In other words, the BWP mechanism allows simple low-bandwidth devices to operate in large 5G channels. LTE is not so flexible: even though the terminals only occupy the PRBs scheduled for them by the eNodeB, they still have to process control signals across the entire channel.

Because of its intrinsic capabilities, in addition to its widespread adoption in unlicensed bands, Wi-Fi will still be a desirable option in certain application scenarios, even after 5G NR [24]. This, together with the fact that many operators have deployed Wi-Fi hotspots for best-effort services to reduce load in congested areas [25], has motivated the 3GPP to standardize 5G support for Wi-Fi [9]. There is thus a clear need for dynamic management capabilities for heterogeneous radio access networks in this context.

### C. MEDIUM SHARING PROCEDURES IN IEEE 802.11 NETWORKS
In contrast to cellular networking transmissions, Wi-Fi transmissions are not scheduled. IEEE 802.11 nodes mitigate collisions with other wireless services nearby through carrier-sense multiple access with collision avoidance (CSMA/CA). The nodes sense the shared channel looking for ongoing communications before starting transmission. If the channel seems to have been idle for some time, the IEEE 802.11 transmitter will send the scheduled frame. Otherwise, it will wait until the existing transmission ends and then sense the channel again.

As a complement to CSMA/CA, IEEE 802.11 devices may also use the optional request to send / clear to send (RTS/CTS) procedure. This mitigates collisions due to the hidden node problem, in which several transmitters that cannot see each other consider that the medium is idle even if some of them are already transmitting. In the RTS/CTS procedure, with a RTS frame, the transmitters notify the receivers and all other nearby nodes of their intention to start a transmission and its duration. The receivers must pick one of the received RTS frames and grant its sender to transmit. To do this, the receivers broadcast CTS frames, which are used not only to notify a transmitter that it is allowed to send a pending frame, but also to inform other IEEE 802.11 devices in the same channel and area about the upcoming transmission and its duration.

Some devices use the CTS-to-self variant of this mechanism to protect transmissions in mixed-mode environments. In this case, the transmitter directly sends a CTS frame and then begins transmission. CTS-to-self is faster and lighter than RTS/CTS, so it may attain higher throughputs. However, it cannot address the hidden node problem.

In order to prevent radar and satellite system disruption in the 5 GHz band, the IEEE 802.11h specification [26] provides the CSA mechanism for IEEE 802.11a and more recent versions to dynamically manage frequency and transmission power. This mechanism was extended to include new capabilities in extended CSA (ECSA) in IEEE 802.11y [27]. Both mechanisms allow Wi-Fi access points (APs) to move BSSs to new frequencies when a radar signal is detected, with minimum station downtime. Without solutions such as CSA or ECSA, AP channel changes cannot be notified to the clients. The channels only become aware of the connection loss when they do not receive any beacons after a timer expires. When this happens, they must discover a new AP by scanning all the channels. They must then authenticate themselves, become associated with the new AP and request an IP address. This would imply a downtime in the order of seconds and the probable loss of any active sessions. With CSA, however, devices are notified about the upcoming channel reconfiguration through CSA information items, which can be included in beacons or sent separately in action frames. These items include the number of target beacon transmit times (TBTT) until the channel switch takes place (0 if the channel switch occurs immediately). Just before their timers expire, the AP and terminals modify their radio interfaces to operate in the new channel.

IEEE 802.11 standards are flexible enough for the manufacturers to apply these mechanisms at will, for example to improve user quality of service (QoS) when the AP detects strong channel interference. An AP that detects significant activity from neighboring BSSs, for example, could move a BSS to a less crowded channel [28].

## III. RELATED WORK

Direct coexistence of Wi-Fi and LTE variants in licensed-exempt bands has been studied in detail [29]–[40]. It has been shown that LTE-U can severely degrade the performance of neighboring Wi-Fi networks as their MAC mechanisms vary considerably [29]–[31]. In [31] the authors analyzed the performance of Wi-Fi coexisting with an LTE-U variant with LBT and found that it was affected more by the presence of LTE nodes than by other Wi-Fi devices.

LAA uses LBT to avoid collisions. Its channel access procedures for transmission resemble Wi-Fi CSMA/CA. However, LAA is unfair to Wi-Fi, since an LAA device is more likely to transmit much longer data bursts and can access the medium much more easily when using transmissions of high-priority classes [33], [34]. This advantage grows with the number of LAA transmitters. In [36] the authors presented an analytical framework based on the well-known Bianchi model [41] to estimate the throughput of Wi-Fi and LAA networks coexisting in the same band. Their model makes some simplifications, such as considering that the backoff counters are not paused when another node begins a transmission. They showed that the aggregated throughput may be higher in scenarios with coexisting IEEE 802.11n and LAA networks with tuned MAC parameters than in homogeneous scenarios with IEEE 802.11n networks. Nonetheless, this higher aggregated throughput is the result of higher LAA throughputs at the expense of less Wi-Fi transmissions. In any case, LAA degrades coexisting Wi-Fi networks significantly. The alternative model presented in [38] for the coexistence between IEEE 802.11n using enhanced distributed channel access (EDCA) and LAA, also based on the Bianchi model, describes the Wi-Fi and LAA backoff procedures more adequately. It considers that each technology and class uses different defer durations and, thus, transmissions of different RATS and classes start decreasing their backoff counters asynchronously. Moreover, it also takes into account that part of the transmitted data may be successfully received when collisions with Wi-Fi transmissions occur. The authors studied coexistence for the full range of EDCA and LAA QoS classes, and the results demonstrated that LAA channel access mechanisms are clearly unfair to Wi-Fi. In [34], [35] the authors also reported that an LAA node that outperforms a Wi-Fi node will interfere with its neighbors more intensely than another Wi-Fi node.

Several other works have improved the fair coexistence of Wi-Fi and LAA networks by dynamically adapting LAA channel access parameters such as contention windows or transmission opportunities, based on channel occupancy time or observed probability of collision [42]–[45]. In [43] the authors compare distributed and cooperative reinforcement learning (RL) algorithms, showing that both fairness and network performance can be further improved when coexisting networks exchange traffic statistics. In [39] it is stated that proportional fairness between LAA and Wi-Fi cannot be guaranteed with standard LAA channel access parameters. However, by dynamically tuning the LAA contention window and the defer duration, based both on Wi-Fi and LAA network information, airtime fairness between these technologies is feasible. In [46]–[49] different game theory and machine learning solutions are applied to select the channels and the LTE-U ON-OFF intervals that maximize the throughput of a cellular technology while protecting collocated Wi-Fi networks.

In any case, the above results can be considered an upper bound on the joint performance of coexisting technologies as there are other effects that were not considered in the publications mentioned. For example, Wi-Fi, LTE-U and LAA clear channel assessment (CCA) thresholds are usually much higher than the sensitivity of the corresponding devices. In other words, they may consider that the channel is idle at interference levels above the minimum power level at which they can demodulate the signals. This issue was irrelevant when Wi-Fi was the dominant solution for unlicensed bands, because if a Wi-Fi transmitter detects a Wi-Fi signal, it considers the channel is occupied even if the signal is below the CCA threshold. However, other works [50]–[53] have shown that this may lead to collisions between LTE and Wi-Fi transmissions, because devices cannot detect ongoing transmissions from other technologies. The works in [8], [54] evaluate the coexistence of Wi-Fi and LAA in a real commercial deployment in the city of Chicago. In particular, in [8] it is shown that Wi-Fi communications experience a considerably higher delay when they coexist with LAA transmissions. Moreover, in [54] the authors found that Wi-Fi and LAA throughputs are reduced by 97% and 35% respectively when the networks coexist rather than operate alone.

One of the proposals to address this issue is to integrate certain Wi-Fi MAC procedures into LTE-U and LAA, specifically RTS/CTS or CTS-to-self, to reduce mutual collisions [50]–[52], [55]. In such a scenario, LAA devices would notify the Wi-Fi stations of their intention to initiate a transmission. Accordingly, even if the LAA signal is below the CCA threshold at the Wi-Fi stations, the devices would not transmit during the time specified in the RTS and CTS frames. An additional Wi-Fi interface is required at the LAA devices to perform CSMA/CA to prevent collisions with Wi-Fi signals below the CCA threshold. By avoiding collisions, the contention windows of the devices of both technologies can be initialized to lower values, providing them with easier access to the medium and therefore allowing higher throughputs. The approaches in [53], [56] advocate for the dynamic modification of CCA thresholds based on the traffic load of the coexisting network. By lowering the CCA thresholds, the devices will be better able to detect ongoing transmissions and thus more collisions will be avoided. However, this will also lead to lower channel usage, as the devices will be more likely to mistakenly consider that the channel is occupied. In this regard, the study in [56] analyzes the optimal CCA threshold values that produce higher throughputs under the assumption that the receivers use interference cancellation techniques and are thus able to decode part of the colliding transmissions. The alternative solution in [57], based on the experimental results of [58], consists of a new LAA channel reservation procedure that is more respectful with Wi-Fi devices.

Recent works have studied the coexistence of NR-U and Wi-Fi in the 6 GHz and millimeter wave bands [59]–[61]. To characterize NR-U, they use standard characteristics and mechanisms for NR licensed operation, along with mechanisms and parameters in 3GPP documents [5] and the regulations for unlicensed operation in the frequency bands they consider. The approach in [61] involves multi-user orthogonal frequency division multiple access (MU OFDMA) to schedule Wi-Fi uplink transmissions. This means that the AP can always schedule the transmissions within a Wi-Fi BSS, avoiding collisions between devices within the same network. Nevertheless, this may only be feasible for IEEE 802.11ax (which introduced multiplexing mechanisms) in the new 6 GHz band, as in other cases the technology must support CSMA/CA.

Regarding Wi-Fi channel sharing mechanisms, bandwidth adaptation attracted interest even before the IEEE introduced newer channel bandwidths in 802.11n to complement the legacy 20 MHz bandwidth [62], as it permits a trade-off between capacity, energy efficiency and coverage. Although it was not initially intended for that purpose, the CSA procedure in IEEE 802.11h can be used to implement this adaptation [63]. Several works have proposed mechanisms based on CSA to dynamically change the center frequency of the channel depending on the interference across the band [28], [64], [65] or to perform seamless handoffs [66]–[69]. However, to the best of our knowledge, nobody has applied CSA channel bandwidth adaptation for dynamic resource sharing between Wi-Fi and other technologies.

In relation to heterogeneous scheduling, centralized coordination of LTE base stations [70]–[73] or Wi-Fi APs [74], [75] has been shown to improve network performance by multiplexing transmissions that might lead to collisions if uncontrolled. In [29], [39], [43], [50] it was also suggested that centralized coordination might enhance performance in a heterogeneous LTE/Wi-Fi scenario. Nevertheless, no mechanisms were proposed in these works for dynamically multiplexing in time transmissions using devices from different technologies. The hybrid base station in [76], [77] serves both Wi-Fi and LAA devices, and the base station in [76] adjusts LAA scheduled resources based on the Wi-Fi load. The proposal in [77] takes advantage of the multi-user MIMO (MU-MIMO) mechanism of IEEE 802.11ax to separate Wi-Fi and LAA transmissions by multiplexing them in space. Nevertheless, this approach is not valid in networks that do not implement this substandard.

Summing up, multiple works have studied the coexistence of IEEE 802.11 and 3GPP standards in unlicensed bands. Nevertheless, taking into account that 3GPP has considered Wi-Fi as an access technology for the newest 5G releases, we propose coordinating Wi-Fi and LTE LAA or NR-U in unlicensed bands to dynamically provide the service levels that heterogeneous clients demand in controlled scenarios. Our main contributions are two solutions for unlicensed spectrum sharing between Wi-Fi and 3GPP technologies that are managed by a single MNO. They involve dynamically multiplexing resources in time or frequency with different levels of granularity, based on the Wi-Fi CTS-to-self and CSA procedures (which were originally intended for different purposes). These proposals are *universal*, that is, they are supported by any current 802.11 COTS end device. We demonstrate that they can also achieve better performance than a scenario in which the two networks directly coexist in the same spectrum. Finally, we provide insights on trade-offs to accommodate varying user traffic demands.

## IV. DYNAMIC RESOURCE SHARING BETWEEN WI-FI AND LTE LAA

Even though 60 GHz bands have ample bandwidth to support extremely high throughput, they experience very intense blockages and therefore frequencies below 6 GHz will still provide advantages in a 5G scenario and beyond. Nevertheless, the resources in such bands are limited. For example, in Europe the unlicensed 5 GHz band only hosts two non-overlapping 160 MHz Wi-Fi channels or six non-overlapping 80 MHz channels. This limits the number of Wi-Fi and LTE LAA channels that an MNO can plan in that band, and their bandwidth.

We envision a scenario in which an MNO, in private facilities (such as industrial ones) or for regulatory reasons, can deploy Wi-Fi or LAA carriers in a set of frequencies without contention or interference with external transmitters. The concept of coordinated channel allocation for orthogonal operation in an unlicensed spectrum is not new in the literature [73], and it has even been considered by the 3GPP, which defined channel access mechanisms for NR-U when the absence of Wi-Fi and other load-based equipment (LBE) devices can be guaranteed on a long-term basis [5], [78].

In this context, an MNO may decide that a Wi-Fi BSS that exclusively occupies a channel will share it with a new LAA cell. The MNO can simply rely on the MAC mechanisms of the transmitters, but by sharing resources with coordinated access it is possible to dynamically distribute the spectrum resources based on current demands and, as we demonstrate in Section VI, improve the overall performance. LTE-LAA uses scheduled channel access whereas Wi-Fi uses random channel access. Thus, the MNO needs mechanisms for multiplexing transmissions from different access networks so that they will share the channel in the proportions (sharing ratio) that are considered optimal at any time.

In the following subsections we consider that an MNO offering data services through an LTE network and a Wi-Fi network wishes to deploy a new LTE LAA carrier. However, as the entire band is occupied in that location, the new carrier will have to share a channel with a pre-existing Wi-Fi BSS. The MNO wishes to maintain tight control over the allocation of channel resources. Because both technologies –Wi-Fi and LTE-LAA– coexist in space, the channel must be shared in time or frequency.

Note that 5G NR-U supports both data aggregation (based on LTE LAA, supplementing the carriers in licensed bands) and standalone operation (based on MulteFire, which relies on LTE LAA) in the unlicensed bands. Moreover, 5G network cores will continue to support LTE RANs [9], meaning that 5G operators may still use LTE LAA to satisfy user demands in unlicensed channels. Of note, our analysis is valid for Wi-Fi and any scheduled wireless technologies such as 5G NR-U, as the legacy Wi-Fi mechanisms that enable both time and frequency dynamic multiplexing are independent of the transmission intervals of the scheduled technologies. In fact, our analysis can be extended to channel sharing between Wi-Fi and *several* other scheduled technologies.

Next we present our two proposals, based on dynamic time and frequency multiplexing, to allow two technologies to share a radio channel.

### A. DYNAMIC TIME MULTIPLEXING APPROACH

In the dynamic time multiplexing (DTM) approach, the channel is alternately allocated to Wi-Fi and LTE LAA transmitters at separate intervals of variable durations. A controller must inform the devices when they are not allowed to use an RAT to avoid interference.

LTE LAA transmissions are scheduled, that is, the terminals transmit (and receive) only when the base station asks them to do so. In contrast, Wi-Fi stations can generally start a transmission at any moment after performing CSMA/CA. Thus, to enable DTM, Wi-Fi APs must use IEEE 802.11 standardized mechanisms that would allow an operator to demand the connected devices to remain silent during the LAA transmission windows, so that the channel can be multiplexed in time.

There are three ways in which an AP can force Wi-Fi end devices to remain silent for a predefined time:

- AP point coordination function (PCF). With PCF, the AP can split Wi-Fi superframes into contention periods (CPs), and contention-free periods (CFPs). The AP can then reserve the channel for the LAA transmitters by avoiding transmissions during all or part of the CFP. However, only one CFP is allowed between any two beacon transmissions. Because a typical beacon interval is 102.4 ms, this solution would force long idle periods between certain transmissions, posing a potential problem, since frames that are ready to be transmitted should be stored until their technology is active. Thus, this approach would impose a long additional delay on data frames (LTE LAA frames in our case). In any case, the IEEE 802.11 group has declared this feature obsolete since the publication of the IEEE 802.11ac standard, and it has been excluded from IEEE 802.11ax.
- Quiet mechanism. This procedure, which was introduced in IEEE 802.11h, is used to force stations to remain silent for a fully configurable interval. However, as in the previous case, the AP can only reserve the channel once per beacon interval, potentially causing long delays.
- CTS-to-self mechanism. This mechanism is used by some IEEE 802.11 transmitters to reserve a channel before initiating a transmission in mixed-mode networks. With this mechanism, the MNO can instruct a Wi-Fi AP to reserve the channel for the LAA transmitters for a given interval. In this case, there are no constraints on the periodicity. Unlike previous mechanisms, the reservation cannot be scheduled in advance. Wi-Fi devices cease operating immediately after receiving the CTS frame.

Therefore, in our context, CTS-to-self is the only feasible solution that can be used by an AP to block Wi-Fi stations with the required flexibility, allowing the channel to be multiplexed in time with transceivers from other wireless technologies. However, the Wi-Fi AP must send a CTS frame just before the beginning of the LTE LAA transmission window.

Figure 1 shows the flow diagram of the DTM approach based on the CTS-to-self mechanism:

1) Based on the loads of the networks and the previous transmission window lengths, the operator determines

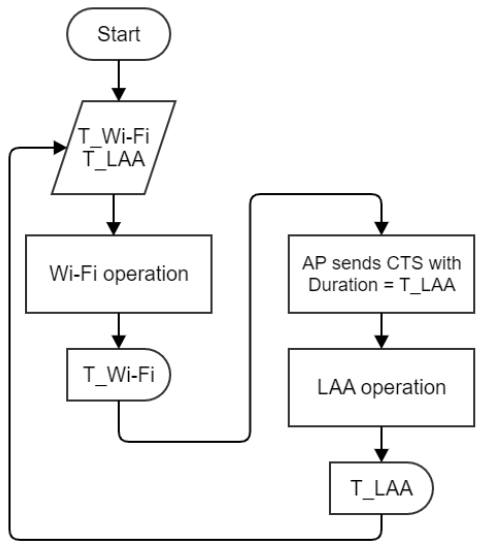

**FIGURE 1.** Dynamic time multiplexing flow diagram.

$T_{Wi-Fi}$ (time reserved for Wi-Fi) and $T_{LAA}$ (time reserved for LAA) for the next interval.
2) The operator assigns the channel to the Wi-Fi BSS devices during $T_{Wi-Fi}$.
3) After $T_{Wi-Fi}$, the channel is allocated to the LTE LAA devices. To this end, the AP asks Wi-Fi stations to refrain from transmitting by sending them a CTS frame, which includes the time the channel will be reserved for LAA $T_{LAA}$. When the Wi-Fi devices receive this frame, they set their network allocation vector (NAV) counter accordingly and wait for it to expire before resuming their backoff countdown.
4) Wi-Fi stations remain silent for at least $T_{LAA}$. During this time, the eNodeB schedules transmissions over the LAA carriers deployed in the channel.
5) After $T_{LAA}$, the NAV counters at the Wi-Fi stations expire, so they resume their backoff windows to contend. Once again, the stations will be allowed to operate for $T_{Wi-Fi}$. Because the eNodeB manages LAA transmissions, LAA interference with Wi-Fi is prevented, as no LAA transmissions are scheduled when the shared channel is not assigned to LTE LAA devices.

Figure 2 shows how the shared channel is multiplexed between Wi-Fi and LTE LAA stations using the DTM approach. The mechanism is highly granular because both the transmission window lengths and their distribution are highly configurable. The standard allows the channel to be reserved for up to 32.767 ms with a granularity of 1 $\mu$s with a single CTS [79]. Longer transmission windows are feasible by concatenating multiple reservations. Consequently, transmission windows can be adapted to the loads of the respective access networks in real time. In other words, depending on the quantity and quality of the service demands of each technology, the MNO can dynamically readjust the transmission periods for the technologies.

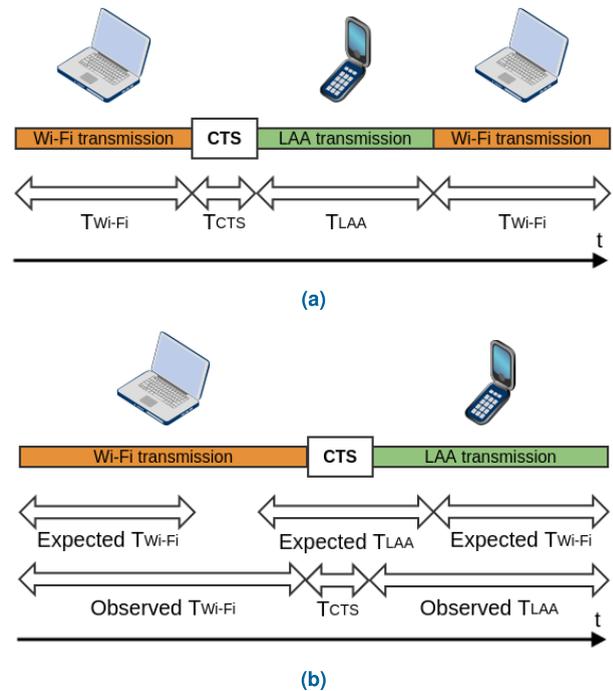

**FIGURE 2.** Dynamic time multiplexing schema. (a) Ideal behavior. (b) The Wi-Fi transmission window is enlarged by an ongoing transmission.

The CTS-to-self mechanism allows the channel to be reserved for adjustable intervals, but the CTS frame has to be transmitted just before the instant when the MNO needs to block Wi-Fi transmissions. However, IEEE 802.11 standards state that in order to avoid collisions the AP should wait for a DCF inter-frame space (DIFS) and start a backoff timer before transmitting the CTS frame. This time could be excessive, so we propose transmitting the CTS frame as soon as the stations are ready to receive it. Even though this proposal would be less desirable in general scenarios, we believe that it is an acceptable simplification in the scenarios we are considering, that is, scenarios where there are strong guarantees that no external wireless services will operate on the channel. Thus, we must wait at least until the active transmitters can change their wireless interfaces from transmission to reception mode. This time is not defined by the standard, but the transmitters can decode acknowledgments (ACKs), which are transmitted a short inter-frame space (SIFS) after the respective data bursts. Accordingly, by waiting for just a SIFS Wi-Fi stations will be successfully able to decode the CTS. In this time gap, the CTS might only collide with the ACK frames. Although these frames can be requested later by the transmitter if they are not able to be sent within the current transmission window, it is more convenient to allow them to be transmitted before reserving the channel for LAA because their retransmission would cause a higher downtime and their loss would require retransmitting the corresponding data frames. However, since the AP takes part in all the communications within the BSS, ACK transmissions can be predicted, meaning the AP can wait until they are transmitted before sending the CTS.

From the perspective of the Wi-Fi AP, uplink transmissions begin at unpredictable times. IEEE 802.11 protocols do not include any preemption mechanism [80] to assist in the transmission of priority frames such as the CTS frame we use to reserve the medium. Thus, Wi-Fi transmission windows may end later than expected to protect ongoing transmissions before reserving the channel for LAA (Figure 2b). In order to achieve the desired sharing ratio between wireless technologies, it may be necessary to increase the duration of certain LTE LAA transmission windows, but this will increase the peak delay of the data frames under high traffic loads.

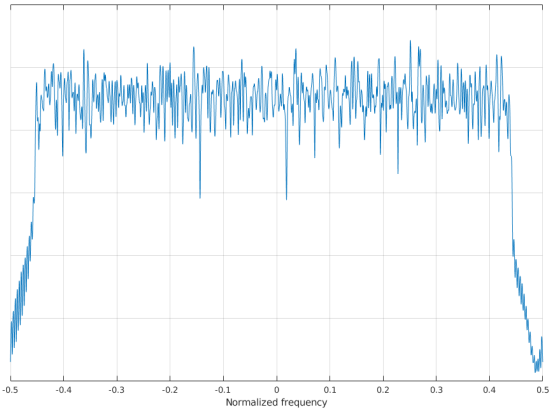

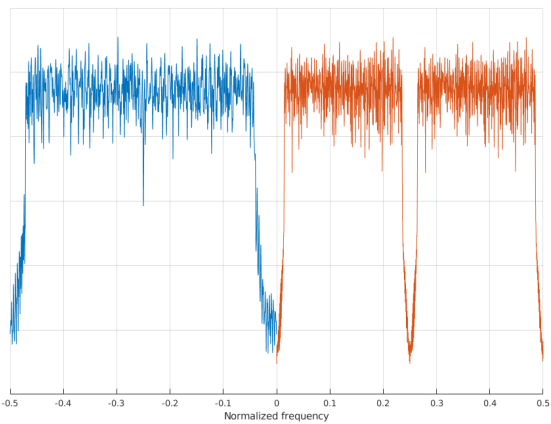

**FIGURE 3.** Dynamic frequency multiplexing schema. (a) Initial situation: the channel is used exclusively by the Wi-Fi BSS. (b) Channel shared by Wi-Fi BSS and LTE LAA carriers.

### B. DYNAMIC FREQUENCY MULTIPLEXING APPROACH

In the dynamic frequency multiplexing (DFM) approach, two technologies can share the radio channel by dividing the spectrum into subbands. The MNO may reduce the bandwidth of the Wi-Fi BSS to introduce new LTE LAA carriers when needed or vice versa. In this setting Wi-Fi and LAA carriers operate simultaneously and independently, as their transmissions are orthogonal in frequency. Figure 3 shows an example of DFM for channel sharing between Wi-Fi and LTE LAA.

If a Wi-Fi AP changes the bandwidth of a BSS channel abruptly, the stations operating in that BSS lose connectivity, as they cannot predict this event. In fact, they only realize that the AP has stopped operating in the original channel after a certain time, which depends on the implementation. Once the connection is lost, the stations must reassociate themselves with the BSS, a process that includes AP discovery, authentication and association and takes seconds. However, as we pointed out in Section II-C, the APs can take advantage of the CSA and ECSA mechanisms to schedule channel switching in such a way that the Wi-Fi stations stay connected to the AP.

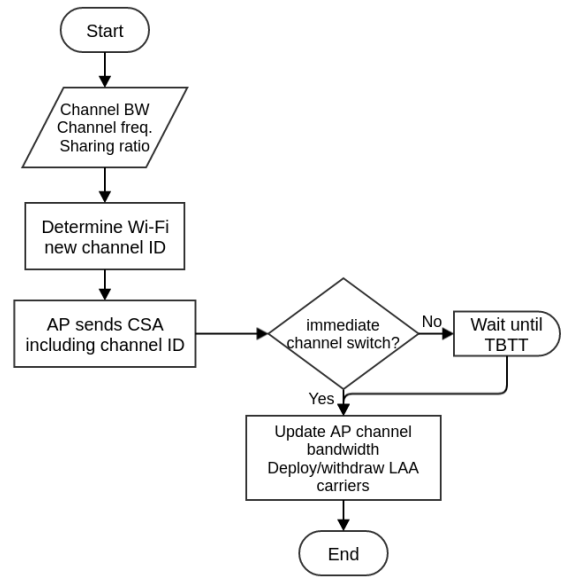

**FIGURE 4.** Dynamic frequency multiplexing flow diagram.

Figure 4 shows the proposed procedure for seamlessly sharing the medium by reducing the Wi-Fi BSS channel bandwidth. The steps are outlined as follows:

1) Once a second technology requests bandwidth, the MNO must decide how to split the Wi-Fi band. To do this, it must determine the ID of the channel to which the BSS needs to be transferred. The channel ID uniquely identifies not only the center frequency of the channel, but also its bandwidth.
2) The Wi-Fi AP must then broadcast an action message including a CSA or ECSA element to inform the stations that the BSS will be moved to the new channel and when this operation takes place. Channel switching can be scheduled to occur at a given TBTT or immediately after the action message.
3) At the channel switching event, the Wi-Fi stations and the AP modify their radio frequency front-ends to operate in the new channel.
4) From the moment channel switching is scheduled, the LAA channels can be deployed in the frequency gap that the AP has freed.

To revert this setup and recover the previous Wi-Fi BSS channel configuration, the MNO simply repeats this procedure by notifying the new channel switching through the Wi-Fi AP.

LAA channel assignments are performed at user level with the radio resource control (RRC) connection configuration protocol [81]. In other words, the cellular network sends an *RRCConnectionReconfiguration* message through the licensed channel to notify each UE which set of unlicensed channels it should operate in. Upon receiving this frame, the UE configures its interface and informs the network via an *RRCConnectionReconfigurationComplete* message. Note that channel assignments are only performed each time the network adds or removes LAA cells with respect to those configured in the UE. These assignments can be performed at any time. Thus, the LAA channel assignment procedure allows for fast, simple and reliable LAA channel resource allocation for DFM.

With DFM, devices from different access technologies can operate simultaneously without interfering with each other. Nevertheless, this approach has a lower sharing granularity than DTM because of the few standard bandwidth partitions defined for both Wi-Fi and LAA. Accordingly, in certain scenarios neither the eNodeBs nor the APs can take advantage of the spectrum that the devices from the respective alternative technologies do not use.

Note that both DTM and DFM require coordination between APs and eNodeBs sharing the channel. This can be achieved in several ways. By following an architecture similar to that described in [72], a centralized orchestrator can decide the allocation of the deployed carriers by notifying the APs and eNodeBs of the bandwidth and intervals in which they can operate (if the carriers belong to different MNOs, a trusted neutral infrastructure provider can operate the orchestrator). An alternative and straightforward approach would be a hybrid base station [76], [77] whereby the coordination entity would be built in the node itself.

## V. ANALYSIS

In this section, we evaluate our proposals to provide insights into their main characteristics, advantages and limitations. First, we analyze their granularity, that is, their degrees of freedom for resource partitioning: valid transmission times in time multiplexing and bandwidth granularity in frequency multiplexing. Then, we evaluate the overhead of the sharing mechanisms. Finally, we analyze the capacity of Wi-Fi and LAA networks using the proposed sharing approaches and when they directly coexist in the medium.

### A. CHANNEL SHARING GRANULARITY

DFM granularity is constrained by the set of standard channel bandwidths (20, 40, 80 and 160 MHz in IEEE 802.11, and 10 and 20 MHz individual channels in LTE LAA, for which up to 80 or 100 MHz bandwidth can be achieved through carrier aggregation [82]). On the one hand, this means that any Wi-Fi BSS reconfigurations leading to bandwidth reductions will divide the channel by at least 2, although complementary narrow-bandwidth BSSs can be introduced to take advantage of the frequency gaps that may appear after reconfigurations. On the other hand, even though LAA and eLAA end devices can be configured for up to 80 MHz aggregated bandwidth in license-exempt bands (100 MHz in the case of feLAA), the network can allocate different subsets of channels to each user and operate within a larger bandwidth. Therefore, while the operator can exploit the entire channel to enhance network capacity, end devices may not be able to reach the peak rates associated with the entire bandwidth allocated to the corresponding technology as they will only be operating in a subset of the frequencies.

In DTM it is not possible to precisely reserve the transmission windows for each technology because the network side cannot govern uplink Wi-Fi transmissions. Transmissions may start at any time if the transmitter wins a contention. Their duration cannot be known in advance. Thus, a DTM transmission window may be enlarged by an ongoing Wi-Fi communication, but its maximum duration can be determined. For example, IEEE 802.11ac sets an upper bound on the duration of very high throughput (VHT) transmissions of 5.484 ms at the PHY layer, regardless of the transmitter's transmission and coding rates [79]. The MAC layer also limits data unit size. As encapsulated data are usually limited to 1,500 B in order to satisfy the maximum transmission unit (MTU) of other network technologies such as Ethernet, Wi-Fi devices frequently use A-MPDU aggregation [79] to build longer transmission bursts. This results in fewer channel access procedures and a higher spectral efficiency [83]. With A-MPDU, multiple IEEE 802.11 MAC frames, also known as MAC protocol data units (MPDUs), are concatenated within a single PHY frame. Up to 64 MPDUs can be aggregated for VHT transmissions, and the maximum A-MPDU-aggregated size is $2^{13+AMPDU_{exp}} - 1$ B, where the A-MPDU length exponent $AMPDU_{exp} \in [0, 7]$ is a parameter reported by Wi-Fi stations to indicate their aggregation capabilities [79]. Thus, the maximum transmission duration of a Wi-Fi device is constrained by both the maximum PHY transmission duration and the maximum A-MPDU length, which, in turn, depends on the A-MPDU capabilities of the station, the physical transmission data rate, and the length of the encapsulated data from the upper stack protocols.

The Wi-Fi transmission window length $T_{Wi-Fi}$ parameter should guarantee the delivery of at least one pre-scheduled transmission within the interval. In other words, a Wi-Fi station should be able to wait for the defer time and its backoff period before starting a transmission. The Wi-Fi defer time or DIFS is 34 $\mu s$. The backoff period consists of a random number of 9 $\mu s$ backoff slots in $[0, CW]$, where $CW$ is the contention window. $CW$ increases after each consecutive collision of the current transmission as $CW = 16 \cdot 2^n - 1$, where $n \in [0, 6]$. Note that the AP knows the backoff period for the next downlink transmission and it may also have information about collisions that have occurred in uplink transmissions within the BSS. This information can be exploited to tune the length of the Wi-Fi transmission window. Moreover, the AP can adapt downlink transmission lengths, by limiting the number of MPDUs that are aggregated in a burst, thereby ensuring that the transmission window does not exceed a predefined target value.

LTE eNodeBs schedule uplink and downlink LAA transmissions. LTE LAA transmissions are composed of 1 ms subframes, each consisting of two 0.5 ms slots. Unlike Wi-Fi transmissions, LAA transmissions are synchronized with the licensed channel and must begin at a slot boundary once the device wins the contention to access the channel. Because of this requirement, LAA transmitters generally occupy the channel until the beginning of the next slot with a reservation signal that does not carry any useful information. In addition, an LAA data burst may end with a partial subframe [84] (that is, the length of the last subframe of the burst is $k \in \{0, 214.29, 428.57, 500, 642.86, 714.29, 785.71, 857.14, 1000\}$ $\mu$s).

The network controller cannot manage the uplink Wi-Fi transmissions. Consequently, the Wi-Fi transmission window lengths cannot be tightly adjusted in advance. At each interval, the operator must set a transmission window length for Wi-Fi operation, $T_{Wi-Fi}$, allowing for the transmission of at least one buffered data burst. The minimum length is:

$$T_{Wi-Fi\_min} = DIFS + CW \cdot 9 \ \mu s. \qquad (1)$$

If a transmission takes place, the window is extended until transmission ends. Thus, a Wi-Fi transmission window can vary according to operator preferences, $T_{Wi-Fi}$, and the maximum transmission window length, $T_{Wi-Fi\_max}$:

$$\begin{aligned} T_{Wi-Fi\_max} = T_{Wi-Fi} + &min(5.484 \text{ ms}, PHY\_header \\ &+ \frac{AMPDU\_length\_max}{data\_rate}) + SIFS \\ &+ T_{Block\_ACK} \end{aligned} \qquad (2)$$

where $AMPDU\_length\_max$ is the length of the aggregated data burst,

$$AMPDU\_length\_max = min(2^{13+AMPDU_{exp}} - 1, 64 \\ \cdot (MPDU\_length + 4)) \text{ B}, \qquad (3)$$

$MPDU\_length$ is the length of the aggregated MPDUs, $PHY\_header$ is the duration of the transmission of the PHY header, $data\_rate$ is the physical data rate for transmitting the data burst and $T_{Block\_ACK}$ is the duration of the transmission of the corresponding block acknowledgment.

LAA transmission window lengths are fully predictable and can be expressed as:

$$T_{LAA} = \Gamma' + n \cdot 0.5 \text{ ms} + k \qquad (4)$$

where $\Gamma'$ is the interval between the start of the transmission window and the next available slot, $n$ is the number of allocated slots within the window and $k$ is the duration of the last partial subframe, if any. The length of LAA transmission windows should be dynamically adjusted based on the real length of the Wi-Fi transmission windows to match the desired sharing ratio between both technologies.

In order to reserve the channel for one of the technologies, the controller must put the other on hold for the desired time. CTS frames allow Wi-Fi transmissions to be held for an adjustable interval with a precision of just 1 $\mu$s. A single CTS frame allows the channel to be reserved for up to 32.767 ms for the LAA operation [79]. Longer LAA transmission windows can be achieved by performing multiple channel reservations. Conversely, channel reservation for Wi-Fi usage has no time constraints, as it only requires LAA data not to be scheduled during the corresponding intervals, irrespective of their length.

### B. EFFECTIVE CHANNEL USAGE

The DFM approach seamlessly isolates transmissions from different technologies, allowing Wi-Fi and LAA devices to operate simultaneously within their respective frequencies without interfering with each other. Each subchannel is permanently available to its transmitters. Therefore, the only reduction in network performance compared to the use of the whole channel is the lower physical peak rate that can be obtained with less bandwidth.

The DTM approach, by contrast, introduces downtimes at each transition from Wi-Fi to LAA operation, as a CTS frame is broadcast to inform the Wi-Fi stations that they must block their transmissions to avoid interfering with LAA devices. The corresponding overhead depends on the transmission windows lengths: the shorter the length, the more CTS frames will be transmitted during a user session. As the eNodeB only needs to avoid scheduling LAA transmissions outside the LAA transmission window to leave the channel vacant, there are no downtimes at the transitions from LAA to Wi-Fi transmission windows.

Since Wi-Fi APs are usually configured to support legacy devices, the CTS frames should be transmitted at the lowest transmission rate using non-High Throughput (non-HT) preambles to ensure they are understood by all connected devices. This also makes CTS frames more robust to interference. CTS transmission time does not depend on the bandwidth of the BSS channel, as different CTSs are simultaneously sent on each of the 20 MHz subchannels of the BSS carrier. The IEEE 802.11ac standard [79] establishes that the CTS transmission should be composed of a PHY preamble and a PHY header, the PLCP service data unit (PSDU) corresponding to the CTS, six zero tail bits and padding bits for the transmission length to match an integer number of OFDM symbols. As we are interested in estimating the total downtime at each transition from Wi-Fi to LAA operation, we must add SIFS to the corresponding transmission times, that is, the minimum time the AP must wait before sending the CTS. Considering the lowest data rate supported at 5 GHz [85], the total downtime during transmission of the CTS is:

$$\begin{aligned} T_{downtime} &= SIFS + T_{CTS \ transmission} \\ &= SIFS + T_{PHY \ preamble} + T_{PHY \ header} + T_{CTS \ PSDU} \\ &\quad + T_{Tail} + T_{Pad \ bits} \\ &= 16 \ \mu s + (10 \cdot 0.8 \ \mu s + 2 \cdot 4 \ \mu s) + 4 \ \mu s \\ &\quad + \left\lceil \frac{(16 \text{ b} + 112 \text{ b} + 6 \text{ b})}{24 \text{ b/symbol}} \right\rceil \cdot 4 \ \mu s \\ &= 60 \ \mu s. \qquad (5) \end{aligned}$$

Effective channel usage with the DTM approach, that is, the proportion of time the channel is available for transmission, is the ratio between the combined length of the Wi-Fi and LTE transmission windows and the combined length plus downtime due to transmission of the CTS. Figure 5 shows the variations in effective channel usage versus the combined length of the Wi-Fi and LTE LAA transmission windows. Effective channel usage is very high, even for short combined lengths (e.g., 99% for a combined transmission length of only 5.940 ms).

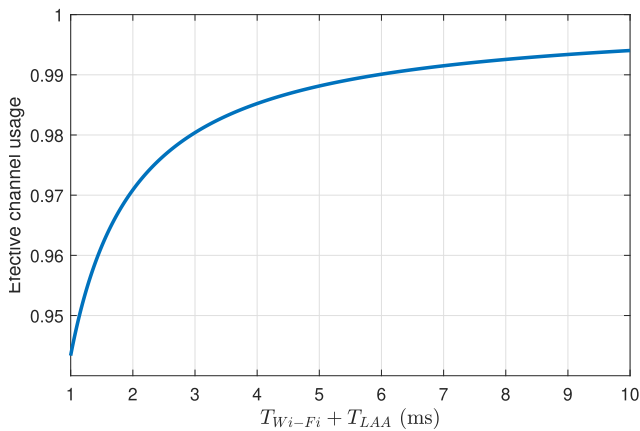

**FIGURE 5.** Dynamic time multiplexing approach: Effective channel usage vs combined length of the Wi-Fi and LTE transmission windows ($T_{Wi-Fi} + T_{LAA}$).

### C. NETWORK CAPACITY

The average network capacity achieved with the DTM approach can be approximated as the capacity that the devices would achieve if they operate in an exclusive channel multiplied by the fraction of time the approach allocates data transmission intervals to their technology. We assume that the devices do not coexist with third technologies during their transmission intervals.

Accordingly, the $C_w^{DTM}$ and $C_l^{DTM}$ capacities of a Wi-Fi network and an LAA network that share a channel using the DTM approach can be approximated as:

$$C_w^{DTM} = C_w^W \frac{\overline{T_{Wi-Fi}}}{\overline{T_{Wi-Fi}} + \overline{T_{LAA}} + T_{downtime}} \quad (6)$$

$$C_l^{DTM} = C_l^W \frac{\overline{T_{LAA}}}{\overline{T_{Wi-Fi}} + \overline{T_{LAA}} + T_{downtime}} \quad (7)$$

$$C_x^W = \left( \left\lfloor \frac{\overline{T_x}}{TXOP_x + \overline{T_{cax}}} \right\rfloor \cdot \frac{TXOP_x + \overline{T_{cax}}}{\overline{T_x}} \right.$$
$$\left. \cdot C_x^{NC}(TXOP_x) \right) + \left( \frac{mod(\frac{\overline{T_x}}{TXOP_x + \overline{T_{cax}}})}{\overline{T_x}} \right.$$
$$\left. \cdot C_x^{NC}\left( max\left(0, mod\left(\frac{\overline{T_x}}{TXOP_x + \overline{T_{cax}}}\right) - \overline{T_{cax}}\right)\right)\right)$$
$$(8)$$

$\overline{T_{Wi-Fi}}$ and $\overline{T_{LAA}}$ are the average durations of the Wi-Fi and LAA transmission windows respectively, and $C_x^W$ in expression (8) is the capacity of technology $x$ ($x = w$ for Wi-Fi and $x = l$ for LAA), considering that downlink transmissions are shortened to fit within the corresponding transmission window. Here, $\overline{T_x}$ is the average transmission window length of technology $x$; $\overline{T_{cax}}$ is the average channel access time that technology $x$ transmitters must wait before they can occupy the medium once the channel is detected as idle when the transmitter is not contending with other peers; $TXOP_x$ is the transmission opportunity for technology $x$ devices for the transmission class that is being evaluated; and $C_x^{NC}(tx\_duration)$ is the capacity that a technology $x$ network would achieve if it did not coexist with other networks within the channel, subject to the constraint that transmissions last for $tx\_duration$. The first term of expression (8) evaluates the contribution of transmissions that achieve maximum $TXOP_x$ duration, whereas the second term considers data bursts that are shorter than $TXOP_x$ at the end of the transmission window (if any). As previously mentioned, uplink Wi-Fi transmissions can enlarge the original Wi-Fi transmission window by using the maximum transmission length defined in the standard. In the downlink, however, the AP is aware of the transmission boundaries, so it can adjust the number of MPDUs it aggregates in each data burst to fit the original transmission window.

**TABLE 1.** Peak physical data rates (Mbps) in IEEE 802.11ac and LTE LAA with carrier aggregation.

|          | 20 MHz | 40 MHz | 60 MHz | 80 MHz | 100 MHz | 160 MHz |
|----------|--------|--------|--------|--------|---------|---------|
| 802.11ac | 86.7   | 200.0  | -      | 433.3  | -       | 866.7   |
| LTE LAA  | 75.4   | 150.8  | 226.1  | 301.5  | 376.9   | -       |

When the channel is shared with the DFM approach, the LAA network capacities mainly depend on the physical rate that the technology can achieve within the allocated bandwidth. This may not be the case for Wi-Fi. As mentioned in Subsection V-A, the restrictions on the amount of data that stations can aggregate or deaggregate may yield shorter transmissions as the bandwidth increases, which in turn leads to more channel contention and, thus, lower proportional capacity. On the other hand, as the channel width increases, the physical Wi-Fi data rate increases nonlinearly, as shown in Table 1. Consequently, depending on the aggregation capabilities of the end devices, the network capacity per MHz may increase or decrease as the channel becomes narrower.

We aim to compare the channel sharing mechanisms proposed in this work with the direct coexistence of Wi-Fi and LAA networks competing for the same channel. To evaluate the network capacities in these cases, we used the analytical model presented in [38] for the downlink. This model analyzes the coexistence of IEEE 802.11n devices with EDCA and LAA devices under saturation conditions and considers that they both generate traffic from their four supported transmission priority classes. We focused, however, on the coexistence of IEEE 802.11ac devices, using the default operation mode and A-MPDU aggregation, with LAA devices that transmit traffic from a single priority class, although the evaluation is then particularized for different LAA classes. The analytical framework in [38] considers very short Wi-Fi transmissions. Accordingly, these can only collide with the first subframe of an LAA burst, or they can even end during the LAA channel reservation procedure, before data transmissions take place. However, when Wi-Fi devices aggregate data frames, this is rarely the case. Thus, we adapted the model to analyze the coexistence of Wi-Fi devices without EDCA with LAA nodes transmitting traffic from a single class. We modified its expressions for the durations of successful and colliding Wi-Fi transmissions, and for the average duration of a contention slot and the throughput of Wi-Fi devices, according to the A-MPDU frame structure and transmission duration. We also adapted the expression of the throughput of the LAA devices to support collisions between Wi-Fi and LAA transmissions that may last longer than a single LAA subframe. Finally, we considered that collisions between Wi-Fi and LAA transmissions may cause the loss of multiple LAA slots: LAA receivers successfully demodulate subframes that do not collide with the Wi-Fi burst (if any). The adapted model is as follows:

$$b_{x,0,0} = \frac{1}{\sum_{r=0}^{M_x} PC_x^r (1 + \frac{2+(1-PB_x)(CW_{x,r}-1)}{2(1-PB_x)})} \quad (9)$$

Equation (9) gives the stationary probability for the state of the Markov chain presented in [38] in which the Wi-Fi or LAA transmitter, after concluding the backoff procedure, is ready to transmit a data burst that did not previously cause a collision (i.e. it is not a retransmission). In this formula, $M_x$ is the maximum number of retransmissions for technology $x$, $PC_x$ is the collision probability of a node using technology $x$, $PB_x$ is the probability that an $x$ transmitter pauses its backoff counter owing to an ongoing transmission, and $CW_{x,r}$ is the contention window of a transmitter using technology $x$ at the retransmission stage $r$. This formula is the result of solving the system of equations for the stationary probabilities of the different states of the Markov chain under the normalization condition. Transmission probability for technology $x$, $\tau_x$, is given by:

$$\tau_x = b_{x,0,0} \sum_{r=0}^{M_x} PC_x^r \quad (10)$$

The collision and backoff countdown blocking probabilities [86] for Wi-Fi and LAA transmitters are given by (11), (12), (13) and (14), respectively:

$$PC_w = 1 - (1-\tau_l)^{n_l}(1-\tau_w)^{n_w-1} \quad (11)$$
$$PC_l = 1 - \left[(1-P_{fc}) + P_{fc}(1-\tau_w)^{n_w}\right](1-\tau_l)^{n_l-1} \quad (12)$$
$$PB_w = 1 - \left[(1-\tau_l)^{n_l}(1-\tau_w)^{n_w-1}\right]^{AIFSN-CCA_{min}+1} \quad (13)$$
$$PB_l = 1 - \left[(1-\tau_w)^{n_w}(1-\tau_l)^{n_l-1}\right]^{m_l-CCA_{min}+1} \quad (14)$$

Parameters $n_w$ and $n_l$ are the respective number of Wi-Fi and LAA transmitters (APs and eNodeBs), $P_{fc}$ is the probability that a collision between a Wi-Fi and an LAA transmission does not only take place within the LAA channel reservation period, but also affects the LAA data transmission, $AIFSN$ and $m_l$ are the number of backoff slots that Wi-Fi and LAA must wait for after their basic defer periods before resuming their backoff counters (2 for Wi-Fi), and $CCA_{min} = min(AIFSN, m_l)$. Note that if the Wi-Fi devices use A-MPDU aggregation and therefore perform long transmissions, $P_{fc} \to 1$.

The probability that no transmissions take place during a backoff slot is:

$$P_{idle} = (1-\tau_w)^{n_w}(1-\tau_l)^{n_l} \quad (15)$$

The probability of a successful transmission for RAT $x$ is:

$$PS_x = n_x \tau_x (1-\tau_x)^{n_x-1}(1-\tau_{\bar{x}})^{n_{\bar{x}}}, \quad (16)$$

where we denote by $\bar{x}$ the alternative technology to the technology $x$ under consideration. The probabilities of Wi-Fi and LAA nodes colliding with one or multiple nodes from their respective RATs are:

$$PC_{ww} = (1-\tau_l)^{n_l}\left[1-(1-\tau_w)^{n_w}-n_w\tau_w(1-\tau_w)^{n_w-1}\right] \quad (17)$$
$$PC_{ll} = (1-\tau_w)^{n_w}\left[1-(1-\tau_l)^{n_l}-n_l\tau_l(1-\tau_l)^{n_l-1}\right] \quad (18)$$

Similarly, the probability of one or multiple Wi-Fi and LAA nodes colliding with each other is:

$$PC_{wl} = \left[1-(1-\tau_w)^{n_w}\right]\left[1-(1-\tau_l)^{n_l}\right] \quad (19)$$

The durations of successful ($s$) and colliding ($c$) transmissions for Wi-Fi are:

$$TS_w = DIFS + T_{phy}$$
$$+ N_w \left(\frac{D_{MPDU} + D_{MAC} + D_{LLC} + D_{Data}}{DR_w}\right)$$
$$T_{tail+pad}^D + SIFS$$
$$+ T_{phy} + \frac{D_{BlockACK}}{BR} + T_{tail+pad}^B \quad (20)$$
$$TC_w = DIFS + T_{phy}$$
$$+ N_w \left(\frac{D_{MPDU} + D_{MAC} + D_{LLC} + D_{Data}}{DR_w}\right)$$
$$+ T_{tail+pad}^D + ACK_{Tout} \quad (21)$$

We consider that the Wi-Fi transmitters perform A-MPDU aggregation. Therefore, a Wi-Fi data burst consists of a PHY preamble and header, a number $N_w$ of MPDU subframes (which in turn are composed of an MPDU delimiter) and MAC and LLC headers and their payload, and ends with a PHY tail and padding bits [79]. $DIFS$ refers to the Wi-Fi DCF inter-frame space, that is, the interval that Wi-Fi devices must wait for before resuming their backoff counters; $T_{phy}$ is the duration of the transmission of the PHY preamble and header; $D_{MPDU}$, $D_{MAC}$, $D_{LLC}$, $D_{Data}$ and $D_{BlockACK}$ are the sizes of the MPDU delimiter, the MAC and LLC headers, the encapsulated data and the block ACK frames, respectively; $BR$ and $DR_w$ are the basic and transmission rates used in Wi-Fi transmissions; $ACK_{Tout}$ is the interval after which the transmitter considers that a collision has occurred when it does not receive an ACK; and $T_{tail+pad}^D$ and $T_{tail+pad}^B$ are the durations of the transmission of the PHY tail plus the corresponding pad bits of the data PSDU and the block ACK, respectively.

For LAA, the duration is the same for colliding and successful transmissions:

$$TS_l = TC_l = \Gamma + TXOP_l, \quad (22)$$

where $\Gamma$ is the average time an LAA transmitter waits for before starting a transmission at the next available slot.

$T_{cs}$ is the average duration of all the events that may occur in the contention scenario, including successful transmission from a Wi-Fi or LAA device, inter-RAT and intra-RAT collisions, and idle backoff slots,

$$T_{cs} = PS_w TS_w + PS_l TS_l + PC_{ww} TC_w + PC_{ll} TC_l \\ + PC_{wl} \max(TC_w, TC_l) + P_{idle}\sigma \quad (23)$$

where $\sigma$ is the duration of a backoff time slot. The average throughput of the Wi-Fi network in the contention scenario is:

$$Th_w = \frac{PS_w N_w D_{Data}}{T_{cs}} \quad (24)$$

For LAA, the average throughput in the same scenario is:

$$Th_l = \frac{13}{14} \frac{DR_l}{T_{cs}} \left[ PS_l \cdot TXOP_l \\ + PC_{wl} \left\lfloor \frac{\max(0, TC_l - TC_w)}{T_{LAAslot}} \right\rfloor T_{LAAslot} \right], \quad (25)$$

where $T_{LAAslot}$ is the duration of the LAA transmission slot. In both cases, the throughput is computed as the quotient between the amount of payload data transmitted during each transmission burst and $T_{cs}$. Note that the LAA frames consist of several slots. Accordingly, when LAA transmissions are longer than Wi-Fi transmissions, the last slots of the LAA data burst may not collide with the Wi-Fi transmission and be received successfully. This is supported by the second term of expression (25). On the contrary, even if the Wi-Fi transmitter performs A-MPDU aggregation and, thus, multiple independent MPDUs are transmitted in a single burst, if a collision at the beginning of a Wi-Fi transmission damages its PHY header, even if none of the aggregated MPDUs collide with the concurring transmission, the receiver will not be able to successfully decode the corresponding frames.

## VI. EVALUATION
### A. ANALYTICAL RESULTS

Considering the analysis of the proposed channel sharing mechanisms in Section V, we evaluate the network capacities in two scenarios: *i)* a scenario that uses standard coexistence mechanisms between Wi-Fi and LAA (direct coexistence) and *ii)* a scenario that uses our proposed mechanisms for dynamic sharing. As in previous research [30], [36], [87] we focus our evaluation on downlink capacity because it consumes most of the channel bandwidth (for example, 3GPP Release 13 only specified downlink operations for LTE-LAA, and Wi-Fi is only used for downlink operations in LWA). The goal is to determine the channel usage distribution that maximizes the aggregated capacity:

$$best_{dma}(bw, ratio) = \underset{dma}{\operatorname{argmax}}\ C_{w+l}(dma, bw, ratio) \quad (26)$$

**TABLE 2.** Wi-Fi parameters for the analytical evaluation.

| Parameter | Value |
|---|---|
| $\sigma$ | 9 $\mu$s |
| $AIFSN$ | 2 |
| $SIFS$ | 16 $\mu$s |
| $DIFS$ | $SIFS + AIFSN \cdot \sigma$ |
| $CWmin_w$ | 16 |
| $CWmax_w$ | 1,024 |
| $M_w$ | 7 |
| $BR$ | 6 Mbps |
| $DR_w$ | See Table 1 |
| $Max\ TXOP_w$ | 5.484 ms |
| $AMPDU_{exp}$ | 7 |
| $N_w$ | 64 |
| $D_{Data}$ | 1,500 B |
| $T_{PHY}$ | 40 $\mu$s |
| $D_{MPDU}$ | 4 B |
| $D_{MAC}$ | 34 B |
| $D_{LLC}$ | 8 B |
| $D_{BlockACK}$ | 32 B |
| $ACK_{Tout}$ | 50 $\mu$s |

By relying on traffic predictions [88], for a given channel bandwidth (*bw*) and a given sharing ratio (*ratio*), it is assumed that the operator is interested in the dynamic multiplexing approach (*dma*) that maximizes the aggregated capacity $C_{w+l} = \alpha C_w + (1-\alpha)C_l$ in (26), where $C_w$ and $C_l$ are the capacities of the Wi-Fi and LAA networks. The definitions of $C_{w+l}^X$, $C_w^X$ and $C_l^X$ follow directly for the scenarios with direct coexistence of technologies (*C*), independent or orthogonal operation (*NC*), dynamic time multiplexing (*DTM*) and dynamic frequency multiplexing (*DFM*), by replacing label $X$ with $C$, $NC$, $DTM$ and $DFM$, respectively. The formulation corresponds to the general case where the networks have different priorities. In the sequel we assume that their priorities are the same and thus $\alpha = 0.5$ and $C_{w+l} = C_w + C_l$.

Hereinafter, when we refer to an LAA network operating on a channel larger than 20 MHz, we actually mean that the MNO is deploying multiple 20 MHz carriers to operate in that whole bandwidth through carrier aggregation. As already mentioned in Subsection II-A, an LAA end device can only operate in 80 MHz of aggregated unlicensed spectrum (100 MHz for feLAA end devices). However, as each device can be configured by the operator with a different subset of carriers, the operator can use the whole channel to boost its network capacity.

**TABLE 3.** LAA parameters for the analytical evaluation.

| Parameter | Value |
|---|---|
| $\sigma$ | 9 $\mu$s |
| $T_f$ | 16 $\mu$s |
| $T_d$ | $T_f + m_l \cdot \sigma$ |
| $DR_l$ | See Table 1 |
| $T_{LAAslot}$ | 0.5 ms |
| $\Gamma$ | $T_{LAAslot}/2$ |
| $LBT\ CCA$ | Type A2 |

**TABLE 4.** LAA class 1 parameters for the analytical evaluation.

| Parameter | Value |
|---|---|
| $m_l$ | 1 |
| $CWmin_l$ | 4 |
| $CWmax_l$ | 16 |
| $M_l$ | 6 |
| $TXOP_l$ | 2 ms |

**TABLE 5.** LAA class 4 parameters for the analytical evaluation.

| Parameter | Value |
|---|---|
| $m_l$ | 7 |
| $CWmin_l$ | 16 |
| $CWmax_l$ | 1,024 |
| $M_l$ | 10 |
| $TXOP_l$ (coexistence) | 8 ms |
| $TXOP_l$ (sharing) | 10 ms |

**TABLE 6.** Wi-Fi capacity in Mbps versus channel bandwidth. No coexistence, $AMPDU_{exp}$ = 7, data payload of 1,500 B.

| | 20 MHz | 40 MHz | 80 MHz | 160 MHz |
|---|---|---|---|---|
| $C_w^{NC}$ | 81.00 | 184.31 | 377.22 | 684.21 |
| $C_w^{NC}$/MHz | 4.05 | 4.61 | 4.72 | 4.28 |

For our analytical evaluation we used the physical data rates (Table 1) and PHY and MAC parameters (Tables 2, 3, 4 and 5) defined in the IEEE 802.11ac [79], [89] and LAA [84], [90] standards. $AMPDU_{exp}$ was set to 7, which is the maximum value supported by the IEEE 802.11ac devices. In the analysis of the DTM approach, for both technologies we established a $TXOP_x$ value to ensure that transmissions fit into the transmission windows. As defined in the 3GPP standards, LAA class 4 transmissions may take up to 10 ms when "the absence of any other technology sharing the carrier can be guaranteed on a long term basis" [90]. In other cases, transmissions should not be longer than 8 ms. Therefore, we set an upper threshold on $TXOP_l$ of 8 ms when LAA class 4 devices coexist with Wi-Fi and 10 ms for all other cases. We adopted the Type A2 channel access procedure for LAA multichannel transmissions [90]. Type A procedures do not experience the additional short delay of Type B access procedures to gain access to the medium once the carrier that is configured as primary occupies the channel. Type A2 procedure uses a single CCA counter for all carriers, whereas Type A1 procedure implements an independent CCA counter for each carrier. Note that because our analysis of the network capacity focuses on downlink traffic, all data transmissions are started by the BSs. Accordingly, the results do not depend on the number of end devices in each network.

In the analytical evaluations that follow we consider one Wi-Fi BSS and one LAA network transmitting downlink traffic by i) sharing the same channel with our dynamic proposals or ii) directly competing in the channel using the default coexistence mechanisms defined by their respective standards. Expressions (9)-(25) were used to compute capacities of the Wi-Fi and LAA networks competing for the medium. The capacity was measured as the downlink throughput under saturation conditions, which is one of the assumptions of the analytical model. These expressions were also used to compute the capacity of orthogonally operating networks. To do this, we set the transmission probability $\tau_x$ and the number of users $n_x$ of the alternative technology to 0 in $Th_w$ and $Th_l$ to obtain $C_w^{NC}$ and $C_l^{NC}$ for the required transmission durations, respectively, which allowed us to compute the capacities under DTM operation with expressions (6)-(7). Thus, under these circumstances, we can use terms "capacity" and "throughput" indifferently ($C_w^{NC} = Th_w$ and $C_l^{NC} = Th_l$). The situation is identical for DFM, where $C_w^{DFM}$ and $C_l^{DFM}$ are also directly $Th_w$ and $Th_l$. In this case, the expressions are evaluated using the physical data rates of the carriers within the bandwidths of each technology. In the sequel, by aggregated capacities in a scenario (direct coexistence or dynamic multiplexing with one of our approaches) we refer to the sum of the Wi-Fi and LAA network capacities in that scenario.

Using the analytical model presented in Subsection V-C we determined the capacities of the Wi-Fi network and the LAA network on a shared 80 MHz channel (Figure 6) for LAA channel access priority classes 1 (Figure 6a) and 4 (Figure 6b) and for a fixed DTM period of 10 ms for the combined duration of the Wi-Fi and LAA transmission windows. Note that because Wi-Fi standards do not consider 60 MHz bandwidth channels, we assume that the operator would deploy two Wi-Fi channels with respective widths of 40 MHz and 20 MHz within the sub-band reserved for this technology for the 75% Wi-Fi - 25% LAA DFM sharing configuration. Otherwise, 20 MHz of this sub-band would be left unused. The operator may use techniques such as those in [66]–[69] to seamlessly balance the load and handoff of Wi-Fi devices considering the new secondary channel. The results indicate that the sharing techniques allow the operator to not only fine-tune the proportions of the channel allocated to each network, but also to achieve a higher aggregated capacity than with direct coexistence.

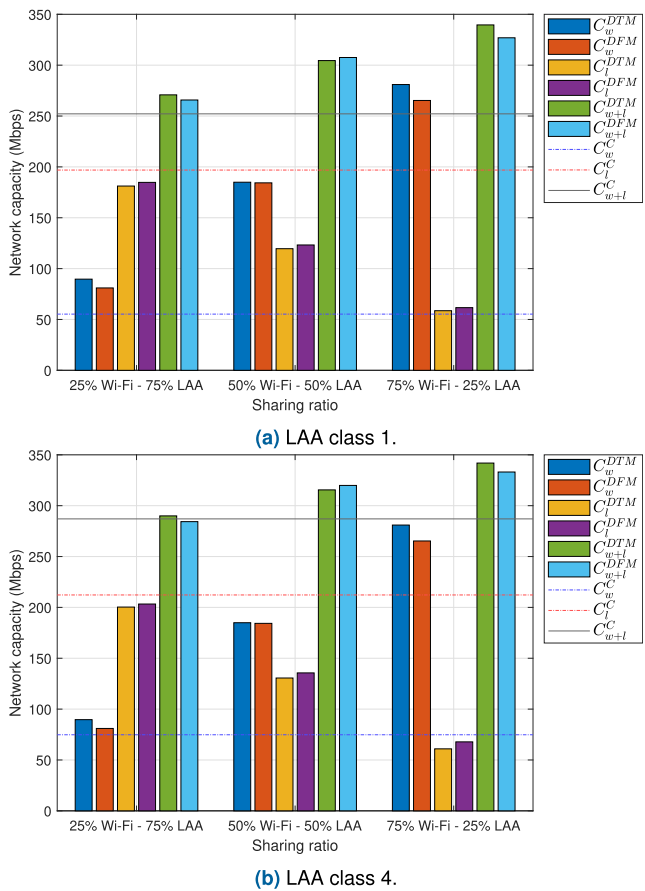

(a) LAA class 1.

(b) LAA class 4.

**FIGURE 6.** Capacities of a Wi-Fi network and an LAA network in an 80 MHz channel. Dynamic time multiplexing (DTM) and dynamic frequency multiplexing (DFM) vs direct coexistence, $T_{Wi-Fi} + T_{LAA} = 10$ ms, $AMPDU_{exp} = 7$, data payload = 1,500 B.

When Wi-Fi and LAA transmitters share an 80 MHz channel, DTM achieves a higher aggregated capacity than DFM in the 25% Wi-Fi - 75% LAA and 75% Wi-Fi - 25% LAA scenarios. In contrast, this capacity is slightly lower in the 50% Wi-Fi - 50% LAA scenario because of the Wi-Fi time and data aggregation constraints defined in IEEE 802.11 standards. The LAA capacity is directly proportional to LAA physical data rates, which, unlike Wi-Fi rates, are proportional to the channel bandwidth. Wider Wi-Fi channels may yield higher physical data rates per MHz, as shown in Table 1. However, Wi-Fi transmissions are bounded in both times and the data length. In this analysis, the payload was set to 1,500 B, resulting in data bursts of up to 96,000 B of data because of the limit of 64 frames that can be acknowledged by a single block ACK, although an A-MPDU length exponent of 7 would support transmission lengths of up to 1,048,575 B. Thus, the wider the Wi-Fi channel, the shorter the Wi-Fi data bursts, resulting in a lower spectrum efficiency as the devices must perform channel access procedures more frequently. Table 6 shows the network capacity per MHz of channel bandwidth for a Wi-Fi BSS that does not coexist with other transmitters, A-MPDU length exponent equal to 7, and MPDU payload of 1,500 B. With this setup, 80 MHz channels are the most efficient and are closely followed by 40 MHz channels. The least efficient option is 20 MHz channels. Since LAA capacities are directly proportional to channel width, DTM achieves better results with configurations in which 20 MHz channels are used for Wi-Fi operating with DFM (25% Wi-Fi - 75% LAA and 75% Wi-Fi - 25% LAA sharing ratios) because the 80 MHz channel width attains much higher spectral efficiency. In contrast, DFM provides slightly better results than DTM in the 50% Wi-Fi - 50% LAA sharing scenario with 80 MHz channels. The spectral efficiencies of the 40 and 80 MHz channels are similar, but the DTM approach introduces a downgrade due to the alternation between technologies. Moreover, transmission windowing may further degrade network capacities, as transmissions may be shorter than they would be without the DTM time constraints. This is discussed further below. In any case, both dynamic multiplexing approaches for resource sharing achieve very similar aggregated capacities for all sharing ratios.

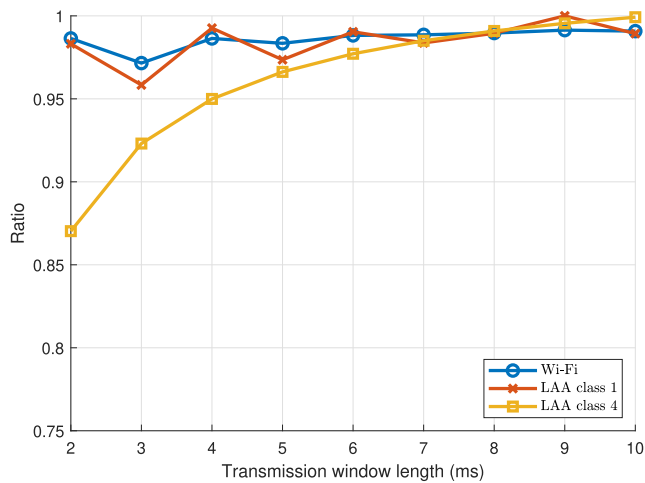

**FIGURE 7.** Ratio between the capacities of Wi-Fi, LAA class 1 and LAA class 4 networks sharing an 80 MHz channel with DTM ($C_x^{DTM}$) and their capacities if DTM windows were enlarged to complete any pending downlink transmissions $\left(C_x^{NC}(TXOP_x) \frac{\overline{T_x}}{\overline{T_x} + \overline{T_{\overline{x}}} + T_{downtime}}\right)$, for different transmission window lengths.

As shown in Subsection V-B, the DTM transmission window lengths affect the network capacity owing to the downtimes produced by the CTS frames that reserve the channel for LAA operation. In addition, operation windowing may shorten data bursts compared to a scenario without time restrictions. This means that more channel-access procedures will be required, which ultimately decreases the capacity of the network. Figure 7 shows the ratio between the capacities of Wi-Fi, LAA class 1 and LAA class 4 networks ($C_x^{DTM}$) for different transmission window lengths, and their capacities if the DTM windows were enlarged to complete any pending transmissions $\left(C_x^{NC}(TXOP_x) \frac{\overline{T_x}}{\overline{T_x} + \overline{T_{\overline{x}}} + T_{downtime}}\right)$.

The LAA networks experience higher performance degradation because the transmissions must start at specific times.

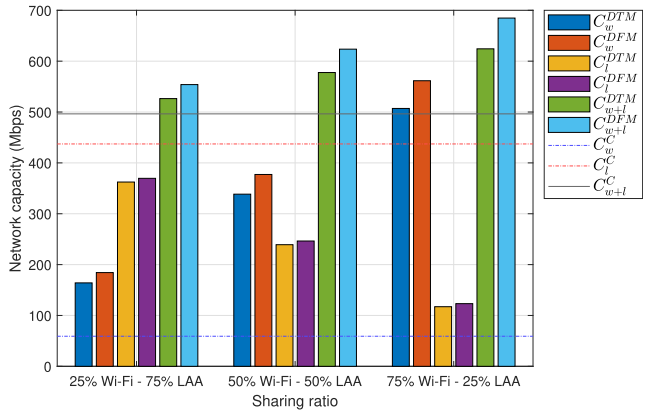

(a) LAA class 1.

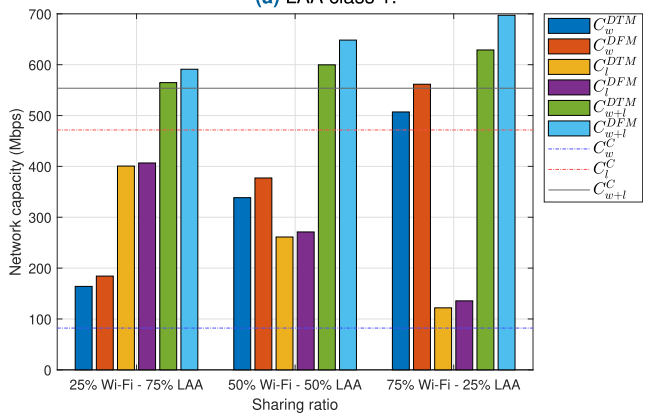

(b) LAA class 4.

**FIGURE 8.** Capacities of a Wi-Fi network and an LAA network in a 160 MHz channel. Dynamic time multiplexing (DTM) and dynamic frequency multiplexing (DFM) vs direct coexistence, $T_{Wi-Fi} + T_{LAA} = 10$ ms, $AMPDU_{exp} = 7$, data payload = 1,500 B.

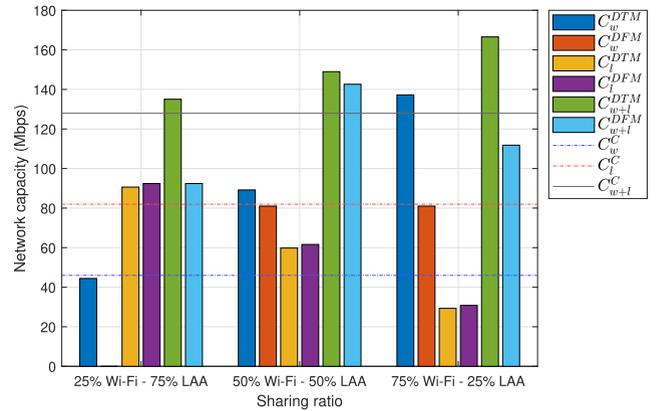

(a) LAA class 1.

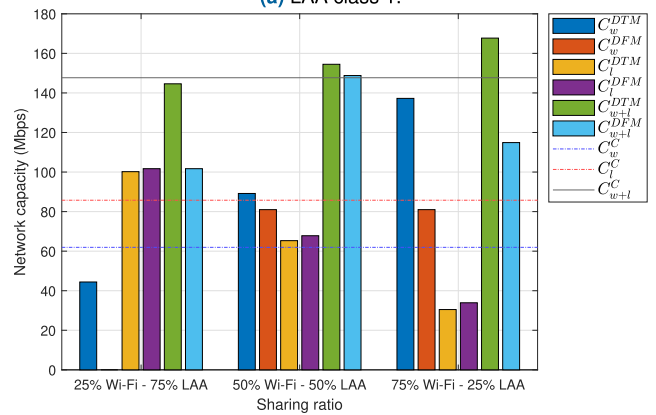

(b) LAA class 4.

**FIGURE 9.** Capacities of a Wi-Fi network and an LAA network in a 40 MHz channel. Dynamic time multiplexing (DTM) and dynamic frequency multiplexing (DFM) vs direct coexistence, $T_{Wi-Fi} + T_{LAA} = 10$ ms, $AMPDU_{exp} = 7$, data payload = 1,500 B.

Accordingly, the transmitters may have to wait until the next LTE subslot boundary. The capacity reduction is greater in LAA class 4 transmissions for DTM transmission windows under 10 ms. Without the constraints imposed by these windows, the transmissions would be much larger and make better use of the spectrum. The degradation due to the confinement of transmissions to their DTM windows, as depicted in Figure 6, may explain the difference when comparing the aggregated capacity achieved by the two proposed dynamic multiplexing approaches. This degradation does not decrease linearly with transmission window enlargement, as it depends on the duration of the transmissions within the window. The operator can optimize the performance by tightly adjusting the transmission window lengths. This is feasible, because Wi-Fi downlink data is buffered for transmission and LAA traffic is scheduled and transmitted in time slots.

Figure 8 shows the capacities of the Wi-Fi network and the LAA network on a shared 160 MHz channel for LAA channel access priority classes 1 (Figure 8a) and 4 (Figure 8b) and for a combined DTM transmission window of 10 ms. As in the evaluation with the 80 MHz channel, two Wi-Fi channels with respective widths of 80 and 40 MHz were aggregated to calculate the capacity in the 75% Wi-Fi - 25% LAA sharing scenario with DFM. In this case, the Wi-Fi 160 MHz channel is less efficient than the 40 and 80 MHz channels, thus DFM achieves higher aggregated network capacities than DTM for all sharing ratios. Nevertheless, both DTM and DFM outperform direct coexistence in terms of aggregated capacity.

Figure 9 shows the capacities of the Wi-Fi network and the LAA network on a shared 40 MHz channel for LAA channel access priority classes 1 (Figure 9a) and 4 (Figure 9b) and for a combined DTM transmission window of 10 ms. In this case, DFM is only worthy in the 50% Wi-Fi - 50% LAA scenario, as Wi-Fi does not allow channels narrower than 20 MHz. Again, the proposed dynamic multiplexing approaches for channel sharing are more advantageous than direct coexistence in terms of aggregated capacity.

Under the conditions so far (typical 1,500 B payload), Wi-Fi data bursts are considerably shorter than LAA ones. The analysis was repeated to evaluate a scenario with Wi-Fi and LAA data bursts of comparable lengths. We used the same parameters shown in tables 2, 3, 4, and 5, but increased the payload length to 15,000 B. Wi-Fi data bursts were considerably longer in this setup, with lengths between those for LAA class 1 and 4 bursts.

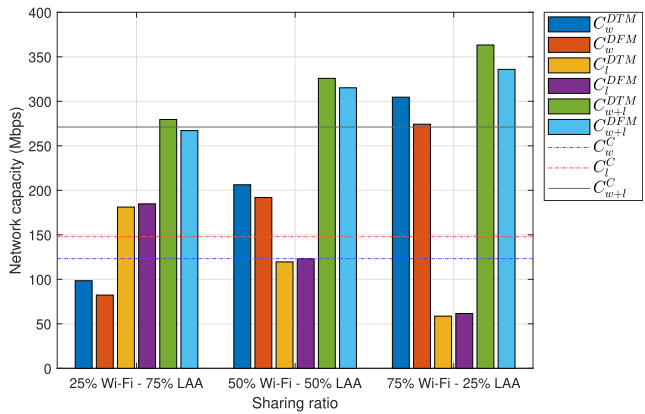

(a) LAA class 1.

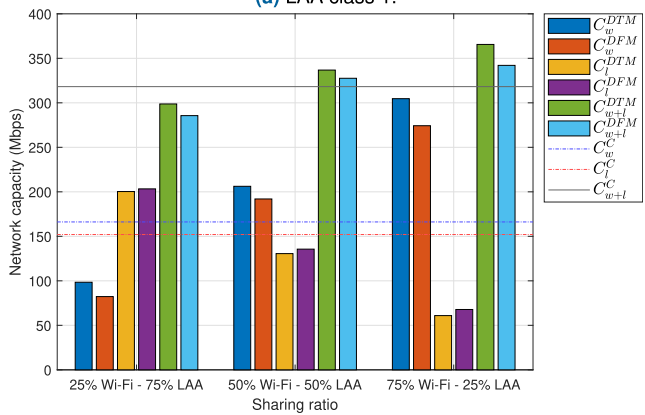

(b) LAA class 4.

**FIGURE 10.** Capacities of a Wi-Fi network and an LAA network sharing an 80 MHz channel. Dynamic time multiplexing (DTM) and dynamic frequency multiplexing (DFM) vs direct coexistence, $T_{Wi-Fi} + T_{LAA} = 10$ ms, $AMPDU_{exp} = 7$, data payload = 15,000 B.

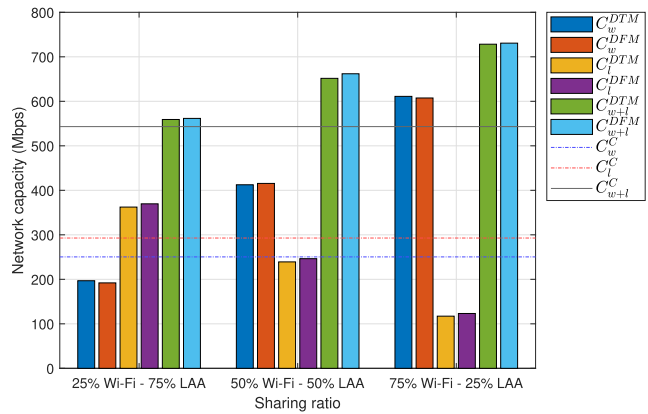

(a) LAA class 1.

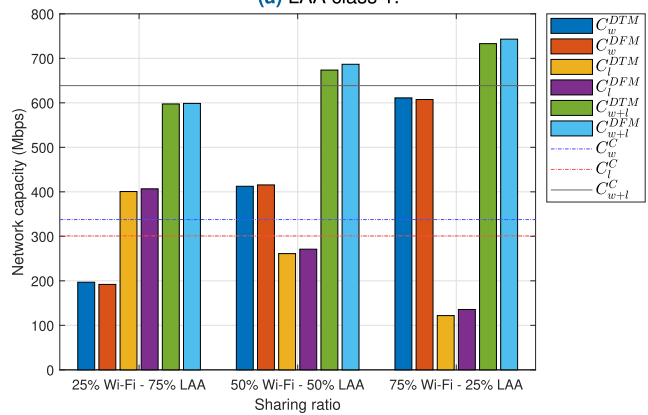

(b) LAA class 4.

**FIGURE 11.** Capacities of a Wi-Fi network and an LAA network sharing a 160 MHz channel. Dynamic time multiplexing (DTM) and dynamic frequency multiplexing (DFM) vs direct coexistence, $T_{Wi-Fi} + T_{LAA} = 10$ ms, $AMPDU_{exp} = 7$, data payload = 15,000 B.

**TABLE 7.** Wi-Fi capacity in Mbps versus channel bandwidth. No coexistence, $AMPDU_{exp} = 7$, data payload of 15,000 B.

|  | 20 MHz | 40 MHz | 80 MHz | 160 MHz |
|---|---|---|---|---|
| $C_w^{NC}$ | 82.30 | 191.98 | 415.51 | 831.92 |
| $C_w^{NC}$/MHz | 4.12 | 4.80 | 5.19 | 5.19 |

Figures 10 and 11 show the network capacities of a Wi-Fi network and an LAA network transmitting downlink traffic with these new settings while sharing the same channel using DTM or DFM or directly coexisting (competing) using standard MAC procedures for LAA channel access priority classes 1 and 4 and for 80 and 160 MHz channel bandwidths, respectively. The combined DTM transmission window is fixed at 10 ms. When they contend, Wi-Fi capacity is significantly higher than in the initial (typical) scenario, and both networks perform similarly. Because they have higher channel access priority, LAA class 1 transmissions attain higher throughput than Wi-Fi transmissions, and similarly, Wi-Fi outperforms LAA class 4. The aggregated capacity is increased by lengthening the Wi-Fi transmissions because this improves the spectrum efficiency. The results also indicate that, even in this second scenario, the proposed dynamic multiplexing approaches for channel sharing outperform direct coexistence in most cases. Direct coexistence is only better in the 25% Wi-Fi - 75% LAA scenario when Wi-Fi coexists with LAA class 4. In contrast, in the case of LAA class 1 transmissions, coexistence is only better than DFM for this sharing ratio in 80 MHz channels.

When Wi-Fi burst durations are not limited by the MAC aggregation constraints defined in the standards, throughputs depend linearly on the physical data rates (see Table 1). This is shown in Table 7. Higher channel bandwidths provide higher network capacities per MHz, except for 160 MHz channels, which yielded the same results as the 80 MHz channels. DTM achieves higher throughput than DFM in 80 MHz channels, as Wi-Fi operation in this bandwidth is much more efficient than in 20 and 40 MHz channels. On the contrary, the difference in efficiency between 160 MHz and 40 and 80 MHz Wi-Fi channels is less than the degradation experienced by the networks with DTM because of the downgrade associated with having to alternate between technologies and the shorter transmission opportunities. As a result, DFM outperformed DTM for all sharing ratios in 160 MHz channels.

Table 8 summarizes the results for the typical scenario with Wi-Fi MPDU payloads of 1,500 B, for all the shared channel bandwidths and sharing ratios in this evaluation. It also highlights the best approach for each configuration according to the utility function (26). As previously stated, DTM outperforms DFM at 40 MHz and for 25% Wi-Fi - 75% LAA and 75% Wi-Fi - 25% LAA sharing ratios in 80 MHz owing to the lower spectral efficiency of the Wi-Fi 20 MHz channels. DFM, however, provides better results at 160 MHz (because of the higher spectral efficiency of 40 and 80 MHz channels) and at 80 MHz for the 50% Wi-Fi-50% LAA sharing ratio (because of the decrease in performance due to the alternation between technologies and the shorter transmission opportunities when using DTM).

**TABLE 8.** Best dynamic multiplexing approach ($best_{dma}$) from expression (26), for class 1 and class 4 LAA networks, for different channel bandwidths (*bw*, in MHz) and sharing ratios (*ratio*). $T_{Wi-Fi} + T_{LAA} = 10$ ms, $AMPDU_{exp} = 7$, data payload = 1,500 B.

| $bw$ | $ratio$ | $C_w$ | $C_l$, class 1 | $C_l$, class 4 | $best_{dma}$ |
|---|---|---|---|---|---|
| 40 | 25% w, 75% l | 44.41 | 90.62 | 100.20 | **DTM** |
| 40 | 50% w, 50% l | 89.17 | 59.81 | 65.32 | **DTM** |
| 40 | 75% w, 25% l | 137.24 | 29.32 | 30.48 | **DTM** |
| 80 | 25% w, 75% l | 89.64 | 181.17 | 200.32 | **DTM** |
| 80 | 50% w, 50% l | 184.31 | 123.24 | 135.60 | **DFM** |
| 80 | 75% w, 25% l | 280.96 | 58.61 | 60.94 | **DTM** |
| 160 | 25% w, 75% l | 184.31 | 369.63 | 406.71 | **DFM** |
| 160 | 50% w, 50% l | 377.22 | 246.39 | 271.11 | **DFM** |
| 160 | 75% w, 25% l | 561.53 | 123.24 | 135.60 | **DFM** |

### B. SIMULATION RESULTS

In order to validate our analytic results we implemented a simulator using the ns-3[1] discrete-event network simulator (the code for our simulation is publicly available at GitHub[2]). The simulator obtains the throughput perceived by a Wi-Fi station and generates a PCAP file with all the data and control messages transmitted by the Wi-Fi devices, proving that LAA channel reservation is performed properly. It fulfils all the practical considerations about granularity and effective channel usage in sections V-A and V-B, without any simplification.

DTM and DFM simulations were performed for different channel bandwidths to evaluate the downlink between a single AP and a single Wi-Fi station and determine the network capacity of a Wi-Fi BSS. Tables 2, 3, 4 and 5 show the parameter values used in the simulations.

Specifically, a single Wi-Fi station was positioned 1 m from the Wi-Fi AP in the infrastructure mode. The AP was directly connected through a point-to-point ideal link to a coordination node that notified in advance the timestamp of the beginning of the next LAA window and its duration. Using this information, the AP adjusted the number of MPDUs to be aggregated in a data burst to expect acknowledgement of the last transmission and also transmit the CTS frame to reserve the channel for LAA operation before the beginning of the LAA transmission window. The CTS included the duration of the LAA transmission window reported by the coordination node in its duration attribute. The AP set its NAV to stop the data and control transmissions during the LAA transmission period. Once the LAA transmission window had ended, the NAV counter expired and the Wi-Fi transmissions were automatically resumed.

The simulation consisted of two stages. In the first stage, the Wi-Fi station was paired with the AP and a `ping` command was issued to populate the address resolution protocol (ARP) tables of both devices. Once these were ready, the AP generated user datagram protocol (UDP) traffic to be transmitted to the station at a higher rate than admissible by the network (i.e., under saturation conditions). This traffic was then transmitted at a rate that the Wi-Fi BSS could manage. The throughput was measured at the application layer for 10 s from the moment of data transmission initiation. Note that throughput values are lower in the simulation than in the evaluation because of the AP control traffic (basically beacon frames).

**TABLE 9.** Capacity of a Wi-Fi network operating in DFM mode for different Wi-Fi allocated channel bandwidths. $AMPDU_{exp} = 7$, data payload = 1,500 B. Simulation vs analytical results.

|  | 20 MHz | 40 MHz | 80 MHz | 160 MHz |
|---|---|---|---|---|
| Analytical | 81.00 | 184.31 | 377.22 | 684.21 |
| Simulation | 79.55 | 181.26 | 372.15 | 677.96 |

Table 9 shows the throughput of a Wi-Fi network for different bandwidths measured using both the analytical model in Section VI and the simulations for different allocated bandwidths. The results are rather similar, validating our model variant for IEEE 802.11ac networks using A-MPDU aggregation. As the analytical model did not consider control frames such as beacons, the analytical results were slightly higher than the simulation results. As such, they can be considered a tight upper bound.

Table 10 compares the analytical and simulation results for a Wi-Fi network operating in DTM mode with a fixed Wi-Fi transmission window of 5 ms for different channel bandwidths and sharing ratios. As in the previous scenario, the simulated throughputs were lower than the analytical throughputs for the same reasons. The analytical results were validated for different transmission window lengths.

Because LAA transmissions are scheduled, the capacity of LAA networks is deterministic when operating on reserved resources. In essence, the throughput depends on the time during which the devices can perform transmissions. One subslot is considered to be misused between every two bursts within the transmission window. As the proposed dynamic multiplexing mechanisms can separate LAA transmissions from Wi-Fi transmissions, and as the transmission window lengths matched their target values during the entire simulation, the analytical throughput of LAA transmissions matched the simulation results.

---

[1]https://www.nsnam.org/
[2]https://github.com/dcandal-gti/Dynamic-Allocation-of-Radio-Resources-to-Wi-Fi-and-Cellular-Technologies-in-Unlicensed-Shared-Freqs

**TABLE 10.** Capacity of a Wi-Fi network operating in DTM mode for a fixed Wi-Fi transmission window of 5 ms and for different channel bandwidths and sharing ratios. $AMPDU_{exp}$ = 7, data payload = 1,500 B. Simulation vs analytical results.

|  | 25% | 50% | 75% |
|---|---|---|---|
| Analytical, 20 MHz | 20.04 | 40.08 | 60.12 |
| Simulation, 20 MHz | 19.57 | 39.29 | 58.93 |
| Analytical, 40 MHz | 44.45 | 88.90 | 133.35 |
| Simulation, 40 MHz | 43.50 | 87.28 | 130.85 |
| Analytical, 80 MHz | 92.19 | 184.38 | 276.57 |
| Simulation, 80 MHz | 90.44 | 181.35 | 271.96 |
| Analytical, 160 MHz | 168.75 | 337.49 | 506.24 |
| Simulation, 160 MHz | 164.80 | 330.74 | 496.26 |

## VII. DISCUSSION

The multiplexing approaches presented in Section IV allow an operator to dynamically assign channel resources to a Wi-Fi network and other scheduled networks without modifications to the end devices. This is convenient, for example, to meet SLAs. In most cases, the proposed multiplexing mechanisms achieve higher throughput than direct coexistence for both typical settings where Wi-Fi transmissions are shorter than LAA transmissions and situations where Wi-Fi data bursts reach the maximum size defined in IEEE 802.11 standards. Note that the analytical model in Subsection V-C does not consider that Wi-Fi transmissions may collide with ongoing LTE-U or LAA transmissions, or vice versa, if the perceived interference is lower than the CCA threshold of the transmitter [50]–[54]. The throughput achieved by Wi-Fi and LAA using standard coexistence methods may thus be even worse in a real scenario, making our proposal even more interesting.

Sharing mechanisms differ in many respects. The DTM approach allows for a wide range of sharing ratios, given the flexibility of channel reservation permitted by the CTS. However, with DFM, these ratios are constrained by standard channel widths. Unlike DFM, for which network behavior is completely deterministic (as devices operate in a subchannel with a fixed channel width), window lengths can be extended with DTM transmission due to unpredictable uplink Wi-Fi transmission events and their durations. In this case, to satisfy the overall sharing ratio, the service provider may need to modify the scheduled length of the next LAA or Wi-Fi transmission window.

In terms of network throughput, different solutions may be preferable depending on the scenario. As shown in Section VI, the Wi-Fi network capacities are not proportional to the channel width. Depending on the data burst lengths, each channel width achieves different proportional capacities, as shown in tables 6 and 7. Table 8 summarizes the analytical results for the typical scenario in which the MPDUs transport a payload of 1,500 B. With this configuration, DTM outperforms DFM for 40 MHz and, in the case of unbalanced sharing ratios, for 80 MHz. Nevertheless, unlike DFM, DTM introduces additional jitter owing to the alternation between technologies.

In this study, we considered that an MNO that serves its devices through both Wi-Fi and LAA access networks operates in a set of frequencies without interference. As mentioned in Section IV, this is a reasonable assumption in scenarios such as industrial facilities. The DTM approach requires CTS messages to preempt other Wi-Fi transmissions. CTS frames are processed not only by the nodes within the operator's BSS, but also by all other Wi-Fi devices operating within the coverage area of the AP and within the same channel. Consequently, the DTM approach should not be applied in a general scenario in which the operator cannot guarantee that no other transmitters will be contending for the channel, because otherwise, all Wi-Fi devices in the area would be forced to transmit within the Wi-Fi transmission window. This is not the case for DFM. Even if an exclusive operation cannot be guaranteed, the operator can dynamically change its BSS bandwidth to deploy or withdraw carriers and ensure that transmissions by different MNOs do not overlap.

Our analysis considers an ideal case where Wi-Fi stations always receive the CTS and CSA messages used to reserve the channel successfully and where no transmitters from third technologies interfere with these messages. In a real scenario, these messages may be lost, and certain Wi-Fi stations may transmit data during LAA transmission intervals. With DTM, some LAA nodes experience transmission outages in slots that the eNodeB allocates to them. With DFM, certain Wi-Fi terminals do not know that the Wi-Fi bandwidth has changed if action frames containing CSA messages are lost. Depending on the implementation, these stations may spend some time unsuccessfully trying to access the previous channel. Because of CTS or CSA frame loss, Wi-Fi devices may use radio resources allocated to LAA devices, but this does not necessarily mean that their transmissions will collide, because Wi-Fi and LAA devices would still use their own MAC procedures. In general, DFM will be more robust than DTM because interference that affects configuration messages will only be harmful just after reconfiguration. For DTM, however, interference may occur each time the channel is allocated to LAA. Nevertheless, the loss of a CSA frame may lead to user session interruption. Therefore, even though the CTS or CSA frame loss does not disable transmissions, it reduces traffic predictability in a controlled scenario. MNOs should consider this when choosing a mechanism. For instance, if an MNO detects many frame retransmissions, even with robust modulations, the DFM approach may be more convenient. If possible, the network core should perform optimal handoffs to assign users to the best available APs.

## VIII. CONCLUSION

Even though a single technology could be used to provide wireless connectivity, there are many situations where it is advantageous to combine multiple RATs. For example, factories could use a scheduled technology to achieve the highest

QoS and Wi-Fi to support legacy devices or best-effort traffic. In addition, even though large portions of spectrum are being freed to deploy 5G NR access networks, MNOs have growing interest in diverting low priority traffic towards unlicensed bands. 5G network cores will support multiple radio access technologies for operation in ISM bands, including LTE LAA, 5G NR-U and Wi-Fi.

Furthermore, 5G aims at supporting multiple use cases with very specific needs in terms of throughput, latency, reliability and energy consumption. This will require dynamic reconfiguration of operator networks so that their resources can be accommodated to sporadic variations in user needs. 5G has introduced new mechanisms to handle this, such as a new core architecture with virtualized network elements, resource slicing, and access networks with bandwidth partitions to dynamically adapt channel resources to active use cases.

To provide MNOs with new tools to manage spectrum in controlled scenarios, in Section IV we have proposed two dynamic multiplexing approaches for jointly deploying Wi-Fi and scheduled technologies such as LTE LAA in a shared channel and for managing the ratio of resources allocated to the different technologies at each moment. These approaches are compatible with current COTS end devices, that is, they do not require any modifications in the terminals. The time-sharing DTM approach in particular is intended for specific scenarios in which the coordinated networks are highly unlikely to coexist with external wireless services, such as in private industrial facilities. This approach avoids competition between technologies sharing the medium by allocating alternate time slots. In other words, any device may use the whole bandwidth of the channel, but only within the transmission window that the MNO allocates to its RAT. The frequency-sharing DFM approach can split the original channel into multiple subchannels at any moment, also ensuring session continuity. This second approach poses no time constraints on the devices, but they can only operate within their allocated bandwidth.

The MNO controls all the LAA and Wi-Fi stations and can allocate channel access to one technology while restricting access to the other technology by placing constraints on time or frequency. One of our main contributions is the definition of mechanisms based on standard Wi-Fi procedures to limit the transmission of COTS Wi-Fi devices.

In Section V we studied when and how the proposals can be used and their associated channel efficiency. An analytical model for the evaluation of the coexistence of an IEEE 802.11ac network using A-MPDU aggregation and an LTE LAA network was presented in Subsection V-C. The analytical results in Subsection VI-A were validated with ns-3 simulations in Subsection VI-B.

We have demonstrated that, in most scenarios, the two sharing methods we propose outperform the direct coexistence of LAA and Wi-Fi networks freely contending for all available spectral resources using their MAC procedures. We have also discussed their respective advantages and disadvantages. DTM allows higher sharing granularity, but DFM is more predictable. The advantage of one mechanism over the other in terms of aggregated throughput depends on the channel bandwidth, the sharing ratios and the aggregation capabilities of the Wi-Fi stations.

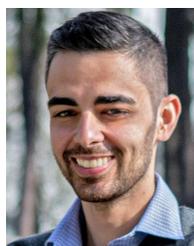

**DAVID CANDAL-VENTUREIRA** received the bachelor's and master's degrees in telecommunication engineering from the University of Vigo, in 2016 and 2018, respectively, where he is currently pursuing the Ph.D. degree in information and communication technologies. Since 2018, he has been working as a Researcher with the Information Technologies Group, University of Vigo. His research interests include mobile and wireless networks and artificial intelligence.

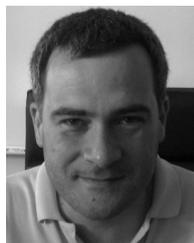

**FRANCISCO JAVIER GONZÁLEZ-CASTAÑO** is currently a catedrático de universidad (Full Professor) with the Department of Telematics Engineering, University of Vigo, Spain, where he leads the Information Technology Group. He has authored over 100 articles in international journals, in the fields of telecommunications and computer science, and has participated in several relevant national and international projects. He holds three U.S. patents.

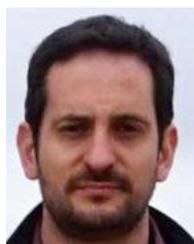

**FELIPE GIL-CASTIÑEIRA** is currently an Associate Professor with the Department of Telematics Engineering, University of Vigo. He has led several national and international research and development projects. He has published over 60 papers in international journals and conference proceedings. He holds two patents in mobile communications. His research interests include wireless communications, core network technologies, multimedia communications, embedded systems, ubiquitous computing, and the Internet of Things.

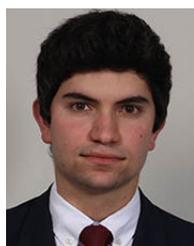

**PABLO FONDO-FERREIRO** received the bachelor's and master's degrees in telecommunication engineering from the University of Vigo, in 2016 and 2018, respectively, where he is currently pursuing the Ph.D. degree. His research interests include SDN, mobile networks, and artificial intelligence. He received the Award for the Best Academic Record from the University of Vigo. He received a Collaboration Grant from the Spanish Ministry of Education for Research on SDN and Energy Efficiency in Communication Networks, in 2016. He received a Fellowship from "la Caixa" Foundation to pursue his Ph.D. degree, in 2018.